\documentclass[reprint,eqsecnum,floats,aps,amsmath,amssymb,nofootinbib,prd,onecolumn, showpacs]{revtex4-1}

\usepackage{graphicx,physics}
\usepackage{amsmath,amssymb,mathtools,mathrsfs}
\usepackage{hyperref}
\usepackage{graphicx}
\usepackage{subfigure}
\usepackage{arydshln}
\usepackage{xcolor,bm}
\usepackage{braket}
\usepackage{tensor}
\usepackage{enumitem,array,textcomp,soul}

\usepackage{color}

\begin{document}
	
	\title{Dapor-Liegener formalism of loop quantum cosmology for Bianchi I spacetimes}
	\author{Alejandro Garc\'ia-Quismondo}
	\email{alejandro.garcia@iem.cfmac.csic.es}
	\affiliation{Instituto de Estructura de la Materia, IEM-CSIC, Serrano 121, 28006 Madrid, Spain}
	\author{Guillermo  A. Mena Marug\'an}
	\email{mena@iem.cfmac.csic.es}
	\affiliation{Instituto de Estructura de la Materia, IEM-CSIC, Serrano 121, 28006 Madrid, Spain}
	
\begin{abstract}
We discuss the quantization of vacuum Bianchi I spacetimes in the modified formalism of loop quantum cosmology recently proposed by Dapor and Liegener. This modification is based on a regularization procedure where both the Euclidean and Lorentzian terms of the Hamiltonian are treated independently. Whereas the Euclidean part has already been dealt with in the literature for Bianchi I spacetimes, the Lorentzian one has not yet been represented quantum mechanically. After a brief review of the quantum kinematics and the quantization of the Euclidean sector, we represent the Lorentzian part of the Hamiltonian constraint by an operator according to the factor ordering rules of the Mart\'in-Benito--Mena Marug\'an--Olmedo prescription. We study the general properties of this quantum operator and the superselection rules derived therefrom, resulting in an action similar to that of the Euclidean operator except that the orientation of the densitized triad is not preserved, a fact which leads to a generic enlargement of the superselection sectors. We conclude with an explanation of the mechanism that prevents this enlargement in the isotropic case and a comment on the effect of alternative prescriptions for the implementation of the improved dynamics.
\end{abstract}

\maketitle
\section{Introduction}

Loop quantum gravity (LQG) stands among the most promising proposals for the formulation of a quantum theory of gravitation \cite{ALQG,Thiem} that would reconcile the principles of the two fundamental pillars of modern theoretical physics: general relativity (GR) \cite{Wald,Hawking-Ellis} and quantum mechanics \cite{CT,Galindo}. It is a nonperturbative quantization of GR independent of any spacetime background structure. In its canonical formulation, the basic variables are holonomies of the Ashtekar-Barbero [$\mathfrak{su}$(2) gauge] connection along closed loops and fluxes of the densitized triad across surfaces. These variables are restricted by a series of constraints that encode the information about the symmetries of the triadic formulation of GR, namely spacetime diffeomorphisms and SU(2) gauge transformations. Following Dirac \cite{Dirac}, these constraints are represented by operators acting on the kinematical Hilbert space of the theory and physical states are required to be annihilated by them.

The techniques of LQG have succeeded in achieving a complete and consistent quantization of cosmological spacetimes. This conjunction of LQG and cosmology led to the birth of a new field of research named loop quantum cosmology (LQC) \cite{AS,LQCG}, that has experienced a rapid evolution in the past years. Within this context, a variety of scenarios have been thoroughly studied, such as Friedmann-Lema\^itre-Robertson-Walker (FLRW) \cite{APS1,APS2,MMO}, Bianchi \cite{chiou1,chiou2,Bianchii,AW-E1,Bianchiii}, and Gowdy \cite{GMM1,GM,Gowdy1,MMW-E} cosmologies. In LQC, the big bang singularity is found to be resolved, which is counted among the most outstanding results in the field. Instead of collapsing, the Universe undergoes a quantum bounce when the energy density becomes comparable with the Planck density. Typically, the spacetime curvature turns out to decrease quickly away from the bounce. Therefore, just a few Planck times after or before the bounce, GR is fit to describe the cosmological dynamics with a very good accuracy. For this reason, it is often said that two \emph{classical} universes (the prebounce one, that contracts, and the postbounce one, that expands) are linked together by the bounce. Moreover, semiclassical states at large volumes have been found to stay peaked across the bounce region and the quantum evolution of their peaks is well approximated by an effective dynamics \cite{APS2}. In this sense, the bounce is referred to as being \emph{deterministic}.

Nonetheless, the formulation of LQC is not unambiguous. Indeed, the regularization of the Hamiltonian constraint (which generates time reparametrizations once the spatial diffeomorphisms and the Gauss constraints have been imposed) involves certain ambiguities \cite{TT,ALM,AAL}, which were originally resolved by appealing to some apparently natural physical criteria. However, a new tendency has arisen recently: other options that result in viable physical pictures are being examined by comparing their physical predictions. This exercise is particularly interesting when it comes to alternatives that lie closer to full LQG, in that their regularization procedure follows more faithfully the strategy of the full theory. These cases have sparked the curiosity of the scientific community lately, resulting in an extensive analysis of the cosmological dynamics and the singularity resolution in these modified alternatives, in order to seek quantitative and/or qualitative differences with respect to the standard LQC picture.

One is led to expect the existence of alternatives that are closer to full LQG because the regularization procedure in LQG and the one that has been commonly employed in LQC have some fundamental differences. These lie in the treatment of the two pieces that compose the gravitational Hamiltonian constraint in GR: the so-called Euclidean and Lorentzian parts. In LQG, each of them requires a separate quantization strategy. The regularization procedure usually employed in LQC, however, relies on the fact that the Euclidean and Lorentzian parts are proportional when the spatial sections of the spacetime under consideration are flat. For this reason, the Hamiltonian constraint has predominantly been quantized as being proportional to the Euclidean part alone. As of yet, we do not fully comprehend the details of how LQG and LQC are related to each other (see, e.g., Refs. \cite{Engle,BK,Engle2,Paw,BEHM}). Therefore, the cosmological dynamics resulting from the standard approach to LQC is not ensured to capture the full cosmological dynamics in LQG. Hence, it seems reasonable to analyze suitable alternative approaches to LQC that are closer to LQG with the objective of shedding light on whether the classical singularity is also resolved by a quantum bounce in the full theory.

Recently, Dapor and Liegener put forward a regularization procedure that was originally conceived in the context of LQG and applied it, without any modification, to LQC \cite{DL1,DL2,DL3}, retrieving an effective Hamiltonian (already considered by Yang, Ding, and Ma \cite{YDM}) that has given rise to a number of works in the past year \cite{Paramc1,Paramc2, genericness,Agullo,Haro,DaPorLie}. Although an examination of the quantum dynamics revealed that the initial singularity was resolved in this formalism, the bouncing picture underwent a qualitative modification. While it still involved a large classical universe, a de Sitter epoch with an emergent Planckian cosmological constant appeared as well. Therefore, the bounce was noted to be asymmetric, such that it either joins a de Sitter contracting solution and a classical expanding universe or the other way around. 

Notably, the asymmetric nature of the bounce appears to be supported by other investigations. For instance, a recent study \cite{gaugeinvariant}, employing for the first time a gauge invariant treatment of the singularity resolution, has reinforced the idea that the bounce needs to be asymmetric. From this perspective, the symmetric bounces obtained in the standard approach to LQC would be an artifact of the gauge fixing in the fluxes of the densitized triad. On the other hand, the authors of Ref. \cite{CSaction} have proposed a procedure to obtain yet another modification of the Hamiltonian constraint, based on a regularization of the Euclidean and Lorentzian parts consisting in expressing them in terms of the Chern-Simons action on the spatial sections. The resulting model (which appears to yield the correct classical limit) is nonsingular, and the big bang is also replaced with a quantum bounce that is asymmetric.

However, all the studies cited above (with the exception of some parts of Ref. \cite{DaPorLie}) have been performed in homogeneous and isotropic spacetimes. For this reason, it is interesting to try and generalize this formalism to less symmetric scenarios such as anisotropic cosmologies. Because of their simplicity, Bianchi I cosmologies (which are spatially flat and homogeneous, but anisotropic \cite{kramer}) appear to be the natural first step. Although these spacetimes have been extensively studied within the framework of LQC, the Lorentzian term of the corresponding Hamiltonian constraint had not been discussed before Ref. \cite{DaPorLie}, where a preliminary analysis of Bianchi I cosmologies was used as the starting point to identify a natural symmetrization that could be respected upon taking the isotropic limit, so as to motivate a preferred factor ordering prescription in flat FLRW spacetimes. In that reference, the Lorentzian part was only regularized classically. Nothing has been said so far about its quantum representation and the subsequent properties of the quantum theory. 

The aim of this paper is precisely to fill this gap in the literature. We will consider Bianchi I cosmologies and discuss their loop quantization in the context of the Dapor-Liegener formalism. In other words, we will regularize the full Hamiltonian constraint without appealing to symmetry considerations and analyze the properties of the quantum theory derived from this modified Hamiltonian. In its quantum representation, we will employ the so-called MMO prescription, which receives its name from the initials of its three authors: Mart\'in-Benito, Mena Marug\'an, and Olmedo. This prescription differs from other existing prescriptions (e.g. the one proposed in Ref. \cite{APS2}) in that it gives rise to some particularly good features in the quantum theory, such as the kinematical decoupling of the quantum analog of the classically singular states and the definition of superselection sectors which are especially simple. It is interesting to see whether those advantages prevail in this scenario.

This paper is structured as follows. In Sec. \ref{QK}, we begin by introducing the quantum kinematical aspects of the model that are relevant to our work. Then, after introducing the Hamiltonian constraint and its different regularizations in Sec. \ref{HC}, we review in Sec. \ref{QEH} the quantum representation of the Euclidean part of the Hamiltonian, focusing on its structure and on the superselection rules defined by its action. Once this task is completed, we proceed to the treatment of the Lorentzian term in Sec. \ref{QLH}. After its quantum representation and the computation of its action, we study in Sec. \ref{sup} whether the superselection sectors of the Euclidean term are altered. We then include a comment on the relation with the superselection rules of the isotropic scenario in Sec. \ref{iso}, and a summary of our results, followed by a brief discussion on the implementation of the improved dynamics  (Sec. \ref{discussion}). Finally, in the Appendix we futher comment on the conventions followed in isotropic and in Bianchi I LQC, and recast the Dapor-Liegener formalism of flat FLRW LQC constructed in Ref. \cite{DaPorLie} in a way such that it is immediately comparable with the present work.

\section{Quantum kinematics}\label{QK}

In this section, we present a brief review of the quantum kinematics of Bianchi I cosmologies \cite{AW-E1,chiou1,Bianchii}. We restrict our analysis to the case where the spatial sections are compact and homeomorphic to a three-torus $T^3$.

The first step in this direction consists in coordinatizing the phase space of the system using Ashtekar-Barbero variables. Unlike in the cases with noncompact spatial sections, the introduction of an auxiliary, finite fiducial cell is not required in principle. Indeed, the model already provides us with a natural choice, namely the entire $T^3$ section, which we take with sides of coordinate length $2\pi$. If we consider a diagonal Euclidean fiducial triad\footnote{As discussed in Ref. \cite{chiou2}, the physical quantities are independent of this choice.}, all the gravitational degrees of freedom are encoded in a pair of canonical variables per orthogonal spatial direction. In more detail, the Ashtekar-Barbero variables can be expressed as
\begin{align}
A_a^i=\dfrac{c^i}{2\pi}\delta_a^i,\quad E^a_i=\dfrac{p_i}{4\pi^2}\delta^a_i,
\end{align}
where $A_a^i$ and $E^a_i$ are the Ashtekar-Barbero connection and the densitized triad, respectively. From now on, spatial indices (ranging from 1 to 3) will be denoted by Latin letters from the beginning of the alphabet ($a,b,c...$). The internal SU(2) indices (labeled by  $i=\theta,\sigma,\delta$), however, will be taken from the middle of the Latin alphabet instead ($i,j,k... $). Furthermore, although we will employ Einstein's summation convention for spatial indices (i.e., whenever two indices are repeated, a sum is to be understood over all the values they can adopt), we will not use it when SU(2) indices are involved.

The connection and triad variables, $c^i$ and $p_i$ respectively, have the following nontrivial Poisson brackets:
\begin{align}
\{c^i,p_j\}=8\pi G\gamma\delta^i_ j,
\end{align}
where $G$ is the Newton constant and $\gamma$ is the Immirzi parameter.

Once the classical phase space of the system has been characterized, we seek its so-called polymeric representation, employed in the implementation of the program of LQC. In order to do so, we first describe the configuration space using holonomies of the Ashtekar-Barbero connection. Actually, given the homogeneity of the model, it suffices to consider holonomies along straight edges in each of the fiducial spatial directions. If we take edges of fiducial length $2\pi\mu_i\in\mathbb{R}$ oriented along the $i$-th direction, we obtain the basic holonomies
\begin{align}
h_{i}^{\mu_i}(c^i)=\text{exp}\left(-\dfrac{i\mu_i c^i\sigma_i}{2}\right)=\cos\left(\dfrac{\mu_ic^i}{2}\right)\mathbb{I}-i\sin\left(\dfrac{\mu_i c^i}{2}\right)\sigma_i,
\end{align} 
where $\mathbb{I}$ is the $2\times 2$ identity matrix and $\sigma_i$ are the Pauli matrices. Finally, we complete the phase space by including the fluxes of the densitized triad through fiducial rectangles that are normal to one of the considered spatial directions. For instance, the triad flux through a rectangle normal to the $i$-th fiducial direction and with fiducial area $S^i$ can be written as $E(S^i)=p_iS^i/(4\pi^2)$.

As usual, the configuration algebra is that of the almost periodic functions of the connection variables, that are generated by the holonomy matrix elements $\prod_ i\mathcal{N}_{\mu_i}(c^i)=\prod_i\text{exp}(i\mu_ic^i/2)$, with $\mu_i\in\mathbb{R}$. Upon quantization, a ket state $\ket{\mu_i}$ will be assigned to each of these complex exponentials. The vector space generated by the linear span of the tensor products of one of such states per spatial direction, $\ket{\mu_\theta,\mu_\sigma,\mu_\delta}=\otimes_{i=\theta,\sigma,\delta}\ket{\mu_i}$, will be denoted by 
\begin{align}
\text{Cyl}_{\text{S}}=\text{span}\{\ket{\mu_\theta,\mu_\sigma,\mu_\delta}:\ \mu_i\in\mathbb{R}\}.
\end{align}
This vector space is the analog of the space of cylindrical functions in full LQG. The kinematical Hilbert space of the system $\mathcal{H}^{\text{kin}}=\otimes_i\mathcal{H}_i^{\text{kin}}$ is obtained by completing the previous vector space Cyl$_{\text{S}}$ with respect to the norm defined by the discrete inner product $\langle \mu_i|\mu_i'\rangle=\delta_{\mu_i,\mu_i{}'}$ in each spatial direction. Therefore, the states $\ket{\mu_\theta,\mu_\sigma,\mu_\delta}$ provide an orthonormal basis of $\mathcal{H}^{\text{kin}}$.

The final step in the characterization of the quantum kinematical aspects of the model consists in the determination of the action of the fundamental operators, namely the triad operator and the basic holonomies, on the kinematical Hilbert space. But, before doing so, a final matter remains to be discussed:  the implementation of the quantization prescription for LQC that is commonly referred to as improved dynamics. 

As it is well-known, in LQG there is a minimum nonzero eigenvalue allowed by the area operator, $\Delta$, which is often called the area gap. Traditionally, one introduces a ``minimum length'' within the framework of LQC in order to account for this feature of the full theory. In practice, one demands that the area of the rectangular holonomy circuits be equal to this area gap. The two different possibilities of imposing this restriction on either the fiducial or the physical area have been studied. The first option leads to the so-called $\mu_0$-scheme, which is known to yield unphysical predictions \cite{APS1}. For this reason, the second option, named the $\bar{\mu}$-scheme or improved dynamics, is preferred in the literature. Notice that, in this latter scheme, the minimum coordinate length is a nontrivial function on phase space, since it depends on the triad variables, which are involved in the expression of the physical area. In the isotropic case, there is a general consensus about how to implement the improved dynamics \cite{APS2}. In anisotropic scenarios, on the other hand, the proposal that is more widely accepted is the one put forward by Ashtekar and Wilson-Ewing in Ref. \cite{AW-E1}. In the following, we will adhere to this proposal, although in Sec. \ref{discussion} we will briefly consider an alternative prescription, presented in Ref. \cite{chiou1}, that leads to a quantum theory which is simpler when the Lorentzian part of the gravitational Hamiltonian is regularized independently. 

According to the proposal of Ashtekar and Wilson-Ewing, the existence of the area gap results in a minimum coordinate length in each fiducial spatial direction $\bar{\mu}_i$ given by
\begin{align}
\bar{\mu}_i=\sqrt{\Delta}\sqrt{\bigg|\dfrac{p_i}{p_j p_k}\bigg|},
\end{align}
where $i$, $j$, and $k$ are understood to be different from one another.

Before writing down the action of the fundamental operators on the orthonormal basis of the kinematical Hilbert space introduced above, it is convenient to relabel the states of this basis in order to simplify the resulting formulae. In the anisotropic case, the optimal choice of parameters is
\begin{align}
\lambda_i (p_i) =\dfrac{\text{sgn}(p_i)\sqrt{|p_i|}}{(4\pi G\gamma\sqrt{\Delta})^{1/3}}.
\end{align}
Using this reparametrization, the action of the basic holonomy operators $\hat{\mathcal{N}}_{\pm\bar{\mu}_i}$ adopts the expression
\begin{align}\label{N}
\hat{\mathcal{N}}_{\pm\bar{\mu}_\theta}\ket{\lambda_\theta,\lambda_\sigma,\lambda_\delta}=\bigg|\lambda_\theta\pm\dfrac{1}{2|\lambda_\sigma\lambda_\delta|},\lambda_\sigma,\lambda_\delta\bigg\rangle,
\end{align}
and similarly for the two remaining spatial directions. Furthermore, these ket states are eigenstates of the triad operators,
\begin{align}
\hat{p}_i\ket{\lambda_\theta,\lambda_\sigma,\lambda_\delta}=(4\pi G\gamma\sqrt{\Delta})^{2/3}\text{sgn}(\lambda_i)|\lambda_i|^2\ket{\lambda_\theta,\lambda_\sigma,\lambda_\delta}.\label{pi}
\end{align}

It is also interesting to discuss the action of the physical volume operator, $\hat{V}=\widehat{\sqrt{|p_\theta p_\sigma p_\delta|}}$. Since it is given by a product of powers of the triad operators, the states of the orthonormal basis introduced in this section are also eigenstates of the volume. For this reason, we may refer to this basis as the ``volume eigenbasis''. From Eq. \eqref{pi}, it follows trivially that the action of the volume operator on the kinematical Hilbert space is given by
\begin{align}
\hat{V}\ket{\lambda_\theta,\lambda_\sigma,\lambda_\delta}=4\pi G\gamma\sqrt{\Delta}|\lambda_\theta\lambda_\sigma\lambda_\delta|\ket{\lambda_\theta,\lambda_\sigma,\lambda_\delta}.
\end{align}
In analogy with the isotropic case, we can define a dimensionless parameter $v$ which is proportional to the physical volume of the Universe. Indeed, in FLRW spacetimes, $V=2\pi G\gamma\sqrt{\Delta}|v|$. Using the same definition, we determine that its anisotropic analog (that we will denote with the same letter for simplicity) takes the value $v=2\lambda_\theta\lambda_\sigma\lambda_\delta$ in the considered basis. 

We can rewrite the action of the fundamental operators in an alternative representation by replacing $\lambda_\theta$ by $v$ (any $\lambda_i$ could be replaced in the same way without any substantial modification, but this choice is the most common one in the literature). The action of the fundamental operators in the $v$-representation ---as opposed to the $\lambda$-representation in Eqs. \eqref{N} and \eqref{pi}--- adopts the form
\begin{align}
\hat{\mathcal{N}}_{\pm\bar{\mu}_\theta}\ket{v,\lambda_\sigma,\lambda_\delta}&=\ket{v\pm\,\text{sgn}(\lambda_\sigma\lambda_\delta),\lambda_\sigma,\lambda_\delta},\\
\hat{\mathcal{N}}_{\pm\bar{\mu}_\sigma}\ket{v,\lambda_\sigma,\lambda_\delta}&=\bigg|v\pm\,\text{sgn}(v\lambda_\sigma),\dfrac{v\pm\,\text{sgn}(v\lambda_\sigma)}{v}\lambda_\sigma,\lambda_\delta\bigg\rangle,\\
\hat{p}_\theta\ket{v,\lambda_\sigma,\lambda_\delta}&=(4\pi G\gamma\sqrt{\Delta})^{2/3}\text{sgn}\left(\dfrac{v}{\lambda_\sigma\lambda_\delta}\right)\dfrac{v^2}{4\lambda_\sigma^2\lambda_\delta^2}\ket{v,\lambda_\sigma,\lambda_\delta},\\
\hat{p}_\sigma\ket{v,\lambda_\sigma,\lambda_\delta}&=(4\pi G\gamma\sqrt{\Delta})^{2/3}\text{sgn}\left(\lambda_\sigma\right)\lambda_\sigma^2\ket{v,\lambda_\sigma,\lambda_\delta}.
\end{align}
The action of the $\delta$-operators is completely analogous to that of the $\sigma$-operators and, thus, we omit it for the sake of brevity.

Throughout the remaining sections, we will employ both the $\lambda$ and the $v$-representations interchangeably, since some results will be more transparent in the former of these representations, and others in the latter.

\section{The Hamiltonian constraint: two regularization procedures}\label{HC}

The fact that GR is a completely constrained theory (i.e., its associated Hamiltonian is an integrated linear combination of constraints) implies that the canonical variables that encode the gravitational degrees of freedom cannot vary arbitrarily but are bound to satisfy a number of constraints instead. In the triadic formulation of GR there is a total of seven such constraints: the three Gauss constraints [these appear as a result of the introduction of an SU(2) gauge symmetry in the triadic formulation], the three momentum or spatial diffeomorphism constraints, and the scalar or Hamiltonian constraint. Since we are considering homogeneous systems, the spatial diffeomorphism constraints are trivially satisfied. Likewise, the Gauss constraints contain no relevant information once a choice of fiducial triad has been made. Hence, the only remaining nontrivial constraint is the Hamiltonian constraint\footnote{In our discussion, we will also refer to it as ``the Hamiltonian''.}. Following Dirac \cite{Dirac}, in the LQG program one seeks to represent this Hamiltonian as an operator acting on the kinematical Hilbert space, and require it to annihilate the physical states. This condition is the quantum mechanical counterpart of the classical vanishing of the constraint.

The gravitational contribution to the Hamiltonian can be written as $H_{gr}(N)=N(H_E+H_L)$, where
\begin{align}\label{HE}
H_E&=\dfrac{1}{16\pi G}\int d^3x\, e^{-1}\sum_ {i,j,k}\epsilon\indices{^{ij}_k}E^a_iE^b_jF\indices{^k_{ab}},\\
\label{HL}
H_L&=-\dfrac{1+\gamma^2}{8\pi G}\int d^3x\,e^{-1}\sum_{i,j}E^a_iE^b_jK^{i}_{[a}K^{j}_{b]}.
\end{align}
In these expressions, $H_E$ and $H_L$ denote the Euclidean and Lorentzian parts of the gravitational Hamiltonian, respectively, and $N$ is the lapse function. Moreover, $e=\sqrt{|\text{det}(E)|}$, $F_{ab}^k$ is the curvature tensor associated with the Ashtekar-Barbero connection $A_a^i$, and $K_a^i$ is the triadic extrinsic curvature. 

In flat and homogeneous LQC, the gauge connection is equal to the triadic extrinsic curvature (up to a constant multiplicative factor, namely, the Immirzi parameter). Furthermore, since the spatial derivatives of the connection must vanish due to homogeneity, the contraction $\sum_k\epsilon\indices{^{ij}_k}F_{ab}^k$ turns out to be proportional to $K_{[a}^iK_{b]}^j$. In other words, the Euclidean and Lorentzian parts are classically proportional in this kind of systems. This observation was crucial in the construction of the standard formalism of LQC. Indeed, the full gravitational Hamiltonian is written solely in terms of the Euclidean part and \emph{then} quantized. That is, a symmetry reduction is performed \emph{before} the regularization and quantization procedures take place. Nevertheless, in principle one cannot ensure that these steps commute. If the Hamiltonian is regularized and quantized in its most general form and the symmetry properties are only applied afterwards, the resulting quantum theory may differ from the one obtained from the aforementioned standard procedure. Recently, there has been a number of works devoted to the study of these differences, both in the full quantum theory and in the effective dynamics derived from it \cite{DL1,DL2,DL3,Paramc1,Paramc2,genericness,Agullo,Haro,DaPorLie}. Nonetheless, all these works focus on flat, homogeneous, and isotropic cosmologies. The aim of this article is to extend this analysis to anisotropic cases. In particular, as we mentioned above, to Bianchi I spacetimes.

Let us begin by writing down the regularized expressions of the Euclidean and Lorentzian parts of the gravitational Hamiltonian in Bianchi I cosmologies. Although the Euclidean part had already been treated in the literature, the Lorentzian one was first discussed in this setting in a recent work of ours (see Ref. \cite{DaPorLie}). Following the procedure described in that reference, we find that $H_E$ and $H_L$ can be written as
\begin{align}
H_E^{\text{BI}}=&\dfrac{1}{16\pi GV}\sum_ip_i\dfrac{\sin \bar{\mu}_ic^i}{\bar{\mu}_i}\sum_{j\neq i}p_j\dfrac{\sin \bar{\mu}_jc^j}{\bar{\mu}_j},\label{HEreg}\\
H_L^{\text{BI}}=&-\dfrac{1}{16\pi GV}\dfrac{1+\gamma^2}{4\gamma^2}\sum_ip_i\dfrac{\sin \bar{\mu}_ic^i}{\bar{\mu}_i}\sum_{k\neq i}\cos\bar{\mu}_kc^k\sum_{j\neq i}p_j\dfrac{\sin \bar{\mu}_jc^j}{\bar{\mu}_j}\sum_{l\neq j}\cos\bar{\mu}_lc^l.\label{HLreg}
\end{align}
In Sec. \ref{QEH}, we will review the quantum representation of the Euclidean Hamiltonian constraint. Then, in Sec. \ref{QLH}, we will proceed to the quantization of the Lorentzian Hamiltonian according to the MMO prescription, which has never been done before.

\section{Quantum Euclidean Hamiltonian}\label{QEH}

The quantum representation of the Euclidean part of the Hamiltonian constraint has already been discussed in a number of works, using different proposals for the implementation of the improved dynamics and different factor ordering prescriptions \cite{chiou1,chiou2,Bianchii,AW-E1,Bianchiii}. In this section, we discuss the quantum representation of Eq. \eqref{HEreg} using the so-called MMO quantization prescription. The motivation for considering this prescription is that, in the isotropic case, it is known to lead to a quantum theory which displays a variety of features that make it simpler and more handleable than the others found in the literature \cite{MMO}. This fact has recently been confirmed as well in the formalism of isotropic LQC based on the independent regularization of the Lorentzian part of the Hamiltonian constraint \cite{DaPorLie}. 

The quantization of the Euclidean constraint using the MMO prescription was first discussed in Ref. \cite{Bianchii}. In this section, we briefly review the most important aspects of such a quantization and refer to the aforementioned article for further details. Following the rules of the MMO prescription (see Sec. IV A of Ref. \cite{DaPorLie}), we perform an algebraic symmetrization of the powers of $\widehat{|p|}$ and $\widehat{1/\sqrt{|p|}}$. Furthermore, we symmetrize the terms involving the sign of the triad variables and the holonomies as
\begin{align}
\text{sgn}(p_i)\sin\bar{\mu}_ic^i\longrightarrow \hat{F}_i=\dfrac{1}{2}\left[\widehat{\text{sgn}(p_i)},\widehat{\sin\bar{\mu}_ic^i}\right]_{+},
\end{align}
where $[\cdot\,,\cdot]_{+}$ denotes the operator anticommutator, and $\widehat{\sin\bar{\mu}_ic^i}=-i(\hat{\mathcal{N}}_{2\bar{\mu}_i}-\hat{\mathcal{N}}_{-2\bar{\mu}_i})/2$.  These rules, applied to Eq. \eqref{HEreg}, lead to an operator on $\mathcal{H}^{\text{kin}}$ of the form
\begin{align}\label{HEop}
\hat{H}_E^{\text{BI}}=\left[\widehat{\dfrac{1}{V}}\right]^{1/2} \left(\sum_{i=\theta,\sigma,\delta}\hat{\mathscr{H}}_E^{(i)}\right) \left[\widehat{\dfrac{1}{V}}\right]^{1/2},
\end{align}
where, for $i\neq j \neq k$, 
\begin{align}
\hat{\mathscr{H}}_E^{(i)}=\dfrac{1}{16\pi G \Delta}\widehat{V}^{1/2}\left(\hat{F}_{j}\hat{V}\hat{F}_k+\hat{F}_k\hat{V}\hat{F}_j\right)\widehat{V}^{1/2}.\label{HEidens}
\end{align}
Let us briefly comment on the factor ordering employed in this definition. First, we notice that the $F$-operators and the volume operator (or, equivalently, any power of it) do not commute. For their products, we have chosen the same ordering as in Ref. \cite{AW-E1}, so that the results can be readily compared\footnote{The specific powers of the volume that appear in Eq. \eqref{HEidens} are due to a particular choice of quantum representation for $1/\bar{\mu}_i$. The operator that one obtains in this way for Bianchi I has an isotropic counterpart that slightly differs in its factor ordering from the conventional operator of flat FLRW. This fact complicates a direct comparison. A detailed discussion of this difference and its consequences is presented in the Appendix.}. The extra symmetrization in Eq. \eqref{HEidens} is due to the fact that $\hat{F}_j$ and $\hat{F}_{k}$ do not commute with each other either, which is a direct consequence of the scheme that we have adopted to implement the improved dynamics. If we had considered instead, e.g., the proposal of Ref. \cite{chiou1}, the minimum coordinate length $\bar{\mu}_i$ in a given spatial direction would only depend on the triad variable in the same spatial direction and, as a result, the commented extra symmetrization would have been unnecessary, with the appropriate factor ordering for the powers of the volume operator \cite{GM}.

The inverse volume operator $\widehat{1/V}$  is defined in the usual manner, by means of a Thiemann identity. One finds that the inverse volume operator annihilates any state that belongs to the kernel of any of the triad operators. We remark that such states are the ones with a vanishing eigenvolume (i.e., the quantum analog of the classical singularities). Hence, by virtue of the algebraic symmetrization that is characteristic of the MMO prescription, the Euclidean part of the quantum Hamiltonian constraint turns out to annihilate the states of zero eigenvolume. Furthermore, it also leaves invariant the orthogonal complement of these singular states, namely the Hilbert space obtained from the completion of $\widetilde{\text{Cyl}}_{\text{S}}=\text{span}\left\{\ket{v,\lambda_\sigma,\lambda_\delta}:\,v\neq 0\right\}$. These two properties allow us to densitize the constraint operator \eqref{HEop}, thereby removing the quantum counterpart of the classical cosmological singularities (that is, the states of vanishing volume). We often say that, in this sense, singularities decouple already at the kinematical level. Formally, this densitization is performed by establishing a bijection between the states $\langle\tilde{\phi}|\in\widetilde{\text{Cyl}}_{\text{S}}^*$ annihilated by the adjoint of $\hat{H}_E^{\text{BI}}$ and the states $\langle\phi |\in\widetilde{\text{Cyl}}_{\text{S}}^*$ annihilated by the adjoint of the \emph{densitized} Hamiltonian $\hat{\mathscr{H}}_E^{\text{BI}}=\sum_i \hat{\mathscr{H}}_E^{(i)}$. From now on, we proceed to study exclusively this densitized version, so that we will actually refer to it even when the denomination ``densitized'' is dropped.

Since the Hamiltonian constraint has a very involved functional form, its spectral properties have not been determined. Nevertheless, it has been shown that it superselects certain Hilbert subspaces (that is, it leaves invariant those subspaces), so that in practice our analysis can be restricted to any of them. Let us now study the superselection sectors of this Hamiltonian operator, for the arguments employed in this simpler case will serve to illustrate our line of reasoning when we deal with the Lorentzian part in Sec. \ref{QLH}. 

\subsection{Superselection sectors}\label{supe}

The superselection rules can be directly derived from the action of $\hat{F}_i$ on the volume eigenbasis. In the $v$-representation,
\begin{align}
\hat{F}_\theta\ket{v,\lambda_\sigma,\lambda_\delta}=&-\dfrac{i}{4}\text{sgn}(\lambda_\sigma\lambda_\delta)\sum_{l=\pm 1}l[\text{sgn}(v)+\text{sgn}(v+2l\,\text{sgn}(\lambda_\sigma\lambda_\delta))]\ket{v+2l\,\text{sgn}(\lambda_\sigma\lambda_\delta),\lambda_\sigma,\lambda_\delta},\label{Ftheta}\\
\hat{F}_\sigma\ket{v,\lambda_\sigma,\lambda_\delta}=&-\dfrac{i}{4}\text{sgn}(\lambda_\sigma)\sum_{l=\pm1}l[1+\text{sgn}(|v|+2l\,\text{sgn}(\lambda_\sigma))]\bigg|v+2l\,\text{sgn}(v\lambda_\sigma),\dfrac{v+2l\,\text{sgn}(v\lambda_\sigma)}{v}\lambda_\sigma,\lambda_\delta\bigg\rangle.\label{Fsigma}
\end{align}
Once again, the action of $\hat{F}_\delta$ is similar to that of $\hat{F}_\sigma$, replacing $\sigma$ with $\delta$.

Already at this level, it is possible to see that, under the action of the Euclidean part of the Hamiltonian constraint, the kinematical Hilbert space splits into eight superselected Hilbert subspaces, with support on each of the eight octants defined by the signs of $v$, $\lambda_\sigma$, and $\lambda_\delta$. Given that each term of the Euclidean part is composed by two $F$-operators corresponding to different spatial directions (up to certain powers of the volume operator that do not significantly affect this discussion\footnote{These additional factors are relevant only when $v=2,4$. In such cases, those powers of the volume operator annihilate the potentially singular contributions arising from the action of $\hat{F}_i$, allowing the well-defined restriction of $\hat{\mathscr{H}}_E$ to the Cauchy completion of $\widetilde{\text{Cyl}}_{\text{S}}$.}), the superselection sectors of $\hat{\mathscr{H}}_E^{\text{BI}}$ can be deduced from those associated with $\hat{F}_i$.

Let us thus analyze the superselection properties of this latter operator. For definiteness, we restrict our analysis to the case where the original state is labeled by three positive numbers, $\ket{\psi}\in\text{Cyl}_{\text{S}}^+=\text{span}\{\ket{v,\lambda_\sigma,\lambda_\delta}:\,v>0,\lambda_r>0\}$, with $r=\sigma,\delta$. Under the action of $\hat{F}_i$, only one of the contributions could potentially change the sign of one (or two, in the case of $\hat{F}_\sigma$ and $\hat{F}_\delta$) of the labels, namely the one corresponding to $l=-1$ in Eqs. \eqref{Ftheta} and \eqref{Fsigma}. The key point is that all these contributions are weighted by a factor that vanishes when $v<2$, which is the only situation in which a sign might change. Indeed, the holonomy operators only produce shifts of two units in $v$ and dilations (or contractions) in the anisotropy variables characterized by a fraction in which the numerator and denominator differ by two units. Furthermore, in the case of  $\hat{F}_\sigma$ and $\hat{F}_\delta$ (which produce both shifts and dilations/contractions), a change in the sign of an anisotropy variable is always bound to a change in the sign of the volume. Therefore, the weighting factor that prevents changes in the sign of the volume $v$ actually prevents any change of sign altogether. In conclusion, we realize that the action of the Euclidean Hamiltonian leaves invariant the Hilbert subspace with support on the octant where all the signs are positive. A similar argument can be applied to the remaining octants. In this way, we conclude that the full kinematical Hilbert space actually splits into eight superselected Hilbert subspaces under the action of $\hat{\mathscr{H}}_{E}^{\text{BI}}$, as we had anticipated.

Actually, we can be more precise in our description of the superselection sectors. Let us focus on the scenario where $\ket{v,\lambda_\sigma,\lambda_\delta}\in\text{Cyl}_{\text{S}}^{+}$. Since each term of the Euclidean part of the Hamiltonian constraint contains two $F$-operators, the constraint produces shifts  in $v$ of four units, if any, and dilations/contractions in one or both anisotropy variables (depending on the term). Hence, the subspaces which are superselected under the repeated action of the Hamiltonian constraint are given by the Cauchy completion of sets of the form 
\begin{align}
\text{Cyl}_{\text{S},\varepsilon,\lambda_\sigma,\lambda_\delta}=\text{span}\{\ket{v^*,\lambda_\sigma^*,\lambda_\delta^*}:\, v^*\in\mathcal{L}^{+}_\varepsilon,\,\lambda_r^*=w_\varepsilon \lambda_r,\,w_\varepsilon\in\mathcal{W}_{\varepsilon}\},
\end{align}	
where $\lambda_r$ are the anisotropy variables of the original state on which the constraint operator repeatedly acts, and $\mathcal{L}^{+}_\varepsilon$ and $\mathcal{W}_{\varepsilon}$ are certain sets defined below. 

On the one hand, it is straightforward to realize that $v^*$ lies on a discrete semilattice of step four, $\mathcal{L}_\varepsilon^+$, defined by a real number $\varepsilon\in(0,4]$, such that
\begin{align}
\mathcal{L}^+_{\varepsilon}=\{v^*=\varepsilon+4n:\, n\in\mathbb{N}\},
\end{align}
where $\mathbb{N}$ denotes the set of nonnegative integers.

On the other hand, the anisotropies of the target states are obtained by the successive multiplication of fractional factors in which the numerator and the denominator differ by two units. Therefore, they always are proportional to the original anisotropies and the proportionality factor $w_\varepsilon$ only depends on $\varepsilon$ [see Eq. \eqref{Fsigma}]. Given these remarks, it is possible to prove that $w_\varepsilon$ can take any value in the discrete set
\begin{align}\label{W+}
\mathcal{W}_{\varepsilon}=\left\{\left(\dfrac{\varepsilon-2}{\varepsilon}\right)^z\prod_{n,m\in\mathbb{N}}\left(\dfrac{\varepsilon+2n}{\varepsilon+2m}\right)^{k^n_m}:\, k_m^n\in\mathbb{N};\ z\in\mathbb{Z}\, \text{ if } \varepsilon>2,\ z=0\ \text{if }\, \varepsilon\leq 2\right\},
\end{align}
which is countably infinite and dense in the positive real line \cite{Gowdy1}. 

In this way, while the volume $v$ has support on discrete semilattices of step four, the anisotropy variables belong to a much more involved set. However, in spite of the obvious complications with respect to the isotropic case, the Euclidean part of the Hamiltonian constraint still superselects separable Hilbert spaces $\mathcal{H}_{\varepsilon,\lambda_\sigma,\lambda_\delta}$, given by the Cauchy completion of $\text{Cyl}_{\text{S},\varepsilon,\lambda_\sigma,\lambda_\delta}$ with respect to the discrete inner product.

\section{Quantum Lorentzian Hamiltonian}\label{QLH}

In this section, we analyze the Lorentzian part of the constraint \eqref{HLreg} and discuss how the results of the previous section are altered when this term is included in the Hamiltonian constraint to be quantized without relating it to the Euclidean contribution by symmetry considerations. 

Let us first discuss the quantum representation of the Lorentzian part following the MMO prescription. We begin by noting that, owing to the factor ordering rules of this prescription, the resulting operator will also be of the form
\begin{align}
\hat{H}_L^{\text{BI}}=\left[\widehat{\dfrac{1}{V}}\right]^{1/2}\hat{\mathscr{H}}_L^{\text{BI}}\left[\widehat{\dfrac{1}{V}}\right]^{1/2},
\end{align}
like in the case of its Euclidean counterpart, and where $\hat{\mathscr{H}}_L^{\text{BI}}$ is the corresponding densitized operator. Then, as it happened with the Euclidean contribution, since the inverse volume operator annihilates the states in the kernel of any of the triad operators and leaves invariant the orthogonal complement obtained by the completion of $\widetilde{\text{Cyl}}_{\text{S}}$, we can adopt a well-defined restriction to the latter, thereby decoupling the quantum states of vanishing volume. In the following we will only study the properties of the densitized counterpart of the Lorentzian Hamiltonian operator.

It is straightforward to realize that the densitized Lorentzian Hamiltonian can be written as 
\begin{align}\label{HLop1}
\hat{\mathscr{H}}_L^{\text{BI}}=-\dfrac{1}{16\pi G}\dfrac{1+\gamma^2}{4\gamma^2}\dfrac{1}{2\Delta}\sum_{i}\sum_{j\neq i}[\hat{\tilde{G}}_i^{(\alpha)},\hat{\tilde{G}}_j^{(\alpha)}]_{+},
\end{align}
where the classical counterpart of $\hat{\tilde{G}}_i^{(\alpha)}$ is\footnote{As we will see, the label $\alpha$ parametrizes the factor ordering adopted for the powers of the volume operator in the quantization.}
\begin{align}
V\text{sgn}(p_i)\sin\bar{\mu}_ic^i\sum_{k\neq i}\cos\bar{\mu}_kc^k.\label{Giclas}
\end{align}
According to the rules of the MMO prescription, we promote the above classical expression to an operator $\hat{\tilde{G}}_i^{(\alpha)}$ on the kinematical Hilbert space in the following manner:
\begin{align}
\hat{\tilde{G}}_i^{(\alpha)}=\dfrac{1}{2}\left[\widehat{\text{sgn}(p_i)},\hat{\tilde{\Theta}}_i^{(\alpha)}\right]_{+},\label{Gia}
\end{align}
where the operator $\hat{\tilde{\Theta}}_i^{(\alpha)}$ is given by 
\begin{align}\label{Thetaia}
\hat{\tilde{\Theta}}_i^{(\alpha)}=\dfrac{1}{2}\hat{V}^{1/2-\alpha}\sum_{k\neq i}\left(\widehat{\sin\bar{\mu}_ic^i}\,\hat{V}^{2\alpha}\widehat{\cos\bar{\mu}_kc^k}+\widehat{\cos\bar{\mu}_kc^k}\,\hat{V}^{2\alpha}\widehat{\sin\bar{\mu}_ic^i}\right)\hat{V}^{1/2-\alpha},
\end{align}
and $\widehat{\cos\bar{\mu}_kc^k}=(\hat{\mathcal{N}}_{2\bar{\mu}_k}+\hat{\mathcal{N}}_{-2\bar{\mu}_k})/2$. In this expression, we restrict to $0<\alpha<1/2$, so that all the considered powers of the volume operator are strictly positive [for negative and vanishing powers, the operator \eqref{Thetaia} would in general become ill defined].

Once the Lorentzian Hamiltonian operator has been introduced, we proceed to the computation of its action on the volume eigenbasis. Given the structure presented above, we begin by discussing the action of $\hat{\tilde{\Theta}}_i^{(\alpha)}$, from which the action of $\hat{\tilde{G}}_i^{(\alpha)}$ can be immediately derived. Let us first consider the operator $\hat{\tilde{\Theta}}_\theta^{(\alpha)}$. In the $v$-representation, 
\begin{align}
2i\,\widehat{\sin\bar{\mu}_\theta c^\theta}\ket{v,\lambda_\sigma,\lambda_\delta}&=\sum_{l=\pm 1}l\,\bigg|{v+2l\,\text{sgn}(\lambda_\sigma\lambda_\delta),\lambda_\sigma,\lambda_\delta}\bigg\rangle,\\
2\,\widehat{\cos\bar{\mu}_\sigma c^\sigma}\ket{v,\lambda_\sigma,\lambda_\delta}&=\sum_{l=\pm 1}\bigg|{v+2l\,\text{sgn}(v\lambda_\sigma),\dfrac{v+2l\,\text{sgn}(v\lambda_\sigma)}{v}\lambda_\sigma,\lambda_\delta}\bigg\rangle.\label{csigma}
\end{align}
The action of the cosine in the $\delta$-direction is entirely analogous to the one presented above. It is straightforward to show that the action of $\hat{\tilde{\Theta}}_\theta^{(\alpha)}$ on $\text{Cyl}_{\text{S}}^{+}$ yields
\begin{align}\label{Thetathetaa}
&\hat{\tilde{\Theta}}_\theta^{(\alpha)}\ket{v,\lambda_\sigma,\lambda_\delta}= -i \dfrac{\pi}{4} G\gamma\sqrt{\Delta}|v|^{1/2-\alpha}\sum_{l,m=\pm1}|v+2l|^{2\alpha}|v+2l+2m\,\text{sgn}(v+2l)|^{1/2-\alpha}\times\nonumber\\
\times&\left\{l\bigg|v+2l+2m\,\text{sgn}(v+2l),\dfrac{v+2l+2m\,\text{sgn}(v+2l)}{v+2l}\lambda_\sigma,\lambda_\delta\bigg\rangle+m\bigg|v+2l+2m\,\text{sgn}(v+2l),\dfrac{v+2l}{v}\lambda_\sigma,\lambda_\delta\bigg\rangle+(\sigma\leftrightarrow\delta)\right\},
\end{align}
which generically comprises a total of 16 states. It is important to note, however, that the states for which $|v+2l|$ or $|v+2l+2m\,\text{sgn}(v+2l)|$ vanish do not contribute. This situation can only occur when $v=2$ or $v=4$. More concretely, we notice that, if $v=2$, the eight states corresponding to $l=-1$ are absent. Something similar happens with the four states corresponding to $l,m=-1$ when $v=4$. We will come back to this observation later in our discussion.

Let us now write down the action of $\hat{\tilde{\Theta}}_\sigma^{(\alpha)}$ for completeness (the action of $\hat{\tilde{\Theta}}_\delta^{(\alpha)}$ is entirely analogous to the one presented below, replacing $\sigma$ with $\delta$). The action of the relevant combinations of holonomy elements goes as follows:
\begin{align}
2i\,\widehat{\sin\bar{\mu}_\sigma c^\sigma} \ket{v,\lambda_\sigma,\lambda_\delta}&=\sum_{l=\pm 1}l\,\bigg|{v+2l\,\text{sgn}(v\lambda_\sigma),\dfrac{v+2l\,\text{sgn}(v\lambda_\sigma)}{v}\lambda_\sigma,\lambda_\delta}\bigg\rangle,\\
2\,\widehat{\cos\bar{\mu}_\theta c^\theta} \ket{v,\lambda_\sigma,\lambda_\delta}&=\sum_{l=\pm 1}\ket{v+2l\,\text{sgn}(\lambda_\sigma\lambda_\delta),\lambda_\sigma,\lambda_\delta},
\end{align}
and the action of the operator $\widehat{\cos\bar{\mu}_\delta c^\delta}$ is obtained from Eq. \eqref{csigma} by replacing $\sigma$ with $\delta$. Particularizing for states that belong to $\text{Cyl}_{\text{S}}^{+}$, we find the following expression for the action of $\hat{\tilde{\Theta}}_\sigma^{(\alpha)}$:
\begin{align}\label{Thetasigmaa}
&\hat{\tilde{\Theta}}_\sigma^{(\alpha)}\ket{v,\lambda_\sigma,\lambda_\delta}= -i\dfrac{\pi}{4} G\gamma\sqrt{\Delta}|v|^{1/2-\alpha}\sum_{l,m=\pm1}|v+2l|^{2\alpha}|v+2l+2m\,\text{sgn}(v+2l)|^{1/2-\alpha}\times\nonumber\\
&\times\left\{l\,\bigg|v+2l+2m\,\text{sgn}(v+2l),\dfrac{v+2l}{v}\lambda_\sigma,\lambda_\delta\bigg\rangle+m\,\bigg|v+2l+2m\,\text{sgn}(v+2l),\dfrac{v+2l+2m\,\text{sgn}(v+2l)}{v+2l}\lambda_\sigma,\lambda_\delta\bigg\rangle\right.\nonumber\\
&+l\,\bigg|v+2l+2m\,\text{sgn}(v+2l),\dfrac{v+2l}{v}\lambda_\sigma,\dfrac{v+2l+2m\,\text{sgn}(v+2l)}{v+2l}\lambda_\delta\bigg\rangle\nonumber\\&\left.+m\,\bigg|v+2l+2m\,\text{sgn}(v+2l),\dfrac{v+2l+2m\,\text{sgn}(v+2l)}{v+2l}\lambda_\sigma,\dfrac{v+2l}{v}\lambda_\delta\bigg\rangle\right\}.
\end{align}
Once again, this formula generically relates $\ket{v,\lambda_\sigma,\lambda_\delta}$ with 16 other states. Nevertheless, as in the case of $\hat{\tilde{\Theta}}_\theta^{(\alpha)}$, certain states do not contribute if $v=2$ or $4$. Indeed, the situation is exactly the same as before, namely the states corresponding to $l=-1$ do not appear if $v=2$, whereas it is only those with both $l$ and $m$ equal to $-1$ that have a vanishing contribution when $v=4$. In this sense, we notice that, in practice, the vanishing of $|v+2l|$ or of $|v+2l+2m\,\text{sgn}(v+2l)|$ restricts the possible values that $l$ and $m$ can adopt in the above sum. 

Therefore, we see that, in the case of the three operators $\hat{\tilde{\Theta}}_i^{(\alpha)}$, not all the states connected with $\ket{v,\lambda_\sigma,\lambda_\delta}\in\text{Cyl}_{\text{S}}^+$ via the consecutive action of two holonomies actually contribute to the final outcome. This is a direct consequence of the factor ordering that we have adopted for the powers of the volume operator. Indeed, the proposed choice precludes the appearance of singular states of any of the two following types: states of zero eigenvolume, on the one hand, and states in which the anisotropies may vanish or diverge, on the other hand. The corresponding restriction on the values of $l$ and $m$ is fundamental to obtain a well-defined densitized Lorentzian Hamiltonian operator on $\mathcal{H}^{\text{kin}}$, as well as to ensure that it admits a proper restriction to the orthogonal complement of the kernel of $\hat{V}$.

An alternative perspective is the following. The commented restriction on the values of $l$ and $m$ in Eqs. \eqref{Thetathetaa} and \eqref{Thetasigmaa} translates into a decoupling of all singular states, both with respect to the volume and the anisotropies, \emph{even at intermediate steps}. In particular, the image states cannot contain contributions with zero eigenvolume, not even after the action of a single holonomy (this is how the terms with quantum number $v^*=v-2$ are removed when $v=2$). In practice, then, a projection $\hat{\mathcal{P}}$ onto the Hilbert completion of $\widetilde{\text{Cyl}_{\text{S}}}$ is performed after the action of each of the holonomy operators that enter the definition of the quantum Lorentzian Hamiltonian. Consequently, we can replace the sine and cosine operators in Eq. \eqref{Thetaia} with their projected counterparts,
\begin{align}\label{proj}
\widehat{\sin\bar{\mu}_ic^i}\longrightarrow \hat{\mathcal{P}}\widehat{\sin\bar{\mu}_ic^i},\qquad \widehat{\cos\bar{\mu}_jc^j}\longrightarrow \hat{\mathcal{P}}\widehat{\cos\bar{\mu}_jc^j}.
\end{align}

Before fixing the value of the parameter of our factor ordering, $\alpha$, it is useful to specify the restrictions on $l$ and $m$ that result from these projections. For this, let us treat $l$ and $m$ as an ordered 2-tuple $(l,m)$ that must belong to a certain set, $I_{v}$, defined as follows for each of the possible values of the volume:
\begin{align}
I_v=\left\{
\begin{array}{ll}
\{(+1,+1),(+1,-1),(-1,+1),(-1,-1)\}&\text{if}\,v\neq 2,4,\\
\{(+1,+1),(+1,-1)\}&\text{if}\, v=2,\\
\{(+1,+1),(+1,-1),(-1,+1)\}&\text{if}\, v=4.
\end{array}
\right.
\end{align}
Then, the restrictions on $l$ and $m$ are easily handled by implementing the replacement $\sum_{l,m=\pm1}\rightarrow\sum_{(l,m)\in I_v}$ in Eqs. \eqref{Thetathetaa} and \eqref{Thetasigmaa}. Note that this replacement  is completely equivalent to introducing the projections \eqref{proj} in the definition of $\hat{\tilde{\Theta}}_i^{(\alpha)}$ [see Eq. \eqref{Thetaia}].

Let us now put forward a concrete choice of factor ordering for the volume operator, i.e. a value for the parameter  $\alpha$. Two options appear as natural. We may either select the totally symmetric ordering ($\alpha=1/4$) or adopt one of the limit values for $\alpha$, namely 0 or 1/2. {In this situation, it seems most reasonable to adopt the value that reproduces the factor ordering that we have presented in Sec. \ref{QEH}, that is the one that has also been adopted in other works on Bianchi I LQC (see Refs. \cite{Bianchii,Bianchiii,AW-E1})}. Comparing the functional form of the densitized Euclidean and Lorentzian parts [see Eqs. \eqref{HEidens} and \eqref{HLop1}], we note that the Euclidean counterpart of the operators $\hat{\tilde{G}}_i^{(\alpha)}$ is actually $\hat{V}^{1/2}\hat{F}_i\hat{V}^{1/2}$. In order to express them in a more similar manner, let us recast $\hat{\tilde{G}}_i^{(\alpha)}$ as
\begin{align}\label{Gia2}
\hat{\tilde{G}}_i^{(\alpha)}=\hat{V}^{1/2-\alpha} \left[\dfrac{1}{4}\widehat{\text{sgn}(p_i)},\sum_{k\neq i}\left( \hat{\mathcal{P}}\widehat{\sin\bar{\mu}_i c^i}\,\hat{V}^{2\alpha}\,\hat{\mathcal{P}}\widehat{\cos\bar{\mu}_kc^k}+\hat{\mathcal{P}}\widehat{\cos\bar{\mu}_kc^k}\,\hat{V}^{2\alpha}\,\hat{\mathcal{P}}\widehat{\sin\bar{\mu}_ic^i}\right)\right]_{+} \hat{V}^{1/2-\alpha}.
\end{align}
It is worth emphasizing that, in this formula, $\widehat{\text{sgn}(p_i)}$ never acts on the kernel of the volume operator, where in principle it would be ill defined, because its action is always preceded either by the volume operator itself or by the projection $\hat{\mathcal{P}}$, that removes that kernel.     

In view of Eq. \eqref{Gia2}, it is inmediate to realize that the two operators $\hat{\Theta}_i$ and $\hat{V}^{1/2}\hat{F}_i\hat{V}^{1/2}$ (for a given $i$) have the same structure if we take the limit $\alpha\rightarrow 0$. We do so respecting the projections $\hat{\mathcal{P}}$ and analyze the resulting operator. Moreover, this factor ordering also has the important advantage that we recover the operator $\widehat{\sin2\bar{\mu}c}$ in isotropic scenarios, a simplification that facilitates the comparison with the results already obtained for FLRW spacetimes in Ref. \cite{DaPorLie} (for more comments on the relation between our construction and the isotropic one, see Sec. \ref{iso} and the Appendix). Let us introduce the following definitions to simplify our notation:
\begin{align}
\hat{\Theta}_i&=\dfrac{1}{2}\left[\hat{\mathcal{P}}\widehat{\sin\bar{\mu}_ic^i},\sum_{k\neq i}\hat{\mathcal{P}}\widehat{\cos\bar{\mu}_kc^k}\right]_{+},\label{Thetai}\\
\hat{G}_i&=\dfrac{1}{2}\left[\widehat{\text{sgn}(p_i)},\hat{\Theta}_i\right]_{+},\label{Gi}\\
\hat{\mathscr{H}}_L^{\text{BI}}&=-\dfrac{1}{32\pi G\Delta}\dfrac{1+\gamma^2}{4\gamma^2}\sum_{i}\sum_{j\neq i}\hat{V}^{1/2}(\hat{G}_i\hat{V}\hat{G}_j+\hat{G}_j\hat{V}\hat{G}_i)\hat{V}^{1/2}.\label{HLop}
\end{align}
If we extract the factors arising from the volume operator, the action of the operators $\hat{\Theta}_i $ can be derived from Eqs. \eqref{Thetathetaa} and \eqref{Thetasigmaa} by taking the limit $\alpha\rightarrow 0$. Indeed, we only need to extract the square roots of  the original physical volume $V=2\pi G\gamma\sqrt{\Delta}|v|$ and of the final volume $2\pi G\gamma\sqrt{\Delta}|v+2l+2m\,\text{sgn}(v+2l)|$. In the same way, one can obtain the action of $\hat{G}_i$ and of $\hat{\mathscr{H}}_L^{\text{BI}}$. However, the resulting number of terms in the action of the Lorentzian part of the Hamiltonian turns out to be considerably large\footnote{Indeed, the action of $\hat{\mathscr{H}}_L^{\text{BI}}$ on $\ket{v,\lambda_\sigma,\lambda_\delta}$ generically leads to a total of 1536 contributions from different states, as opposed to only 24 states in the Euclidean sector.}. For this reason, we are not going to specify its explicit expression, as the authors of Ref. \cite{Bianchii} did with the Euclidean contribution. 

Nonetheless, the complexity of the operator does not prevent us from analyzing the structure of the superselection sectors and extracting physically relevant information. We proceed to discuss in the first place whether the eight octants decouple under its action, as it was the case when considering the Euclidean part alone. Given that the states of the orthonormal basis that we are using are eigenstates of the volume operator, we can determine whether the octants are superselected by examining the properties of the $G$-operators.

As a result of the structure of $\hat{G}_i$ [see Eq. \eqref{Gi}], the action of these operators on the volume eigenbasis takes the form
\begin{align}
\hat{G}_i\ket{v,\lambda_\sigma,\lambda_\delta}=\dfrac{1}{2}\sum_{v^*,\lambda_\sigma^*,\lambda_\delta^*}\left[\text{sgn}(\lambda_i)+\text{sgn}(\lambda_i^*)\right]\ket{v^*,\lambda_\sigma^*,\lambda_\delta^*},
\end{align}
where $v^*$, $\lambda_\sigma^*$, and $\lambda_\delta^*$ are the quantum numbers of the target states obtained from $\ket{v,\lambda_\sigma,\lambda_\delta}$ through the action of $\hat{\Theta}_i$. Therefore, the sum is to be understood over all such states, bearing in mind that $\lambda_\theta=v/(2\lambda_\sigma\lambda_\delta)$. It is important to remark that the coefficient weighting each contribution in the above expression ensures that the states where the orientation of the triad is reversed do not appear (or, more rigorously, the states where the signs of the triad variable in the corresponding spatial direction changes). This is obviously a consequence of the symmetrization of the sign of the triad variables and the holonomy elements that characterizes the MMO prescription. 

We begin by discussing the case of $\hat{G}_\theta$. Disregarding the powers of the volume in Eq. \eqref{Thetathetaa}, we arrive at the following action for $\hat{\Theta}_\theta$ on $\text{Cyl}_{\text{S}}^+$:
\begin{align}\label{Thetatheta}
\hat{\Theta}_\theta\ket{v,\lambda_\sigma,\lambda_\delta}=-\dfrac{i}{8}\sum_{(l,m)\in I_v}
&\left\{l\bigg|v+2l+2m\,\text{sgn}(v+2l),\dfrac{v+2l+2m\,\text{sgn}(v+2l)}{v+2l}\lambda_\sigma,\lambda_\delta\bigg\rangle\right.\nonumber\\
&+\left.m\bigg|v+2l+2m\,\text{sgn}(v+2l),\dfrac{v+2l}{v}\lambda_\sigma,\lambda_\delta\bigg\rangle+(\sigma\leftrightarrow\delta)\right\}.
\end{align}
As far as the volume $v$ is concerned, the target states are related to the original state by shifts of four units, if any. If we take into account that each term of the Lorentzian Hamiltonian contains two $G$-operators and, therefore, two $\Theta$-operators, it is immediate to conclude that $\hat{\mathscr{H}}_L^{\text{BI}}$ connects states of volume $v$ with states of volume $v$ or $v\pm8$. Thus, the action of the full Hamiltonian obviously leaves invariant discrete lattices of step four. However, as we discussed in the previous section, the superselection sectors defined by the action of the Euclidean part are even simpler: the semilattices of positive and negative volume actually decouple. Let us now determine whether this decoupling occurs in the Lorentzian sector as well.

We recall that, in the case of the Euclidean part discussed in Sec. \ref{QEH}, each state obtained by the action of $\hat{F}_i$ is weighted by a factor $[\text{sgn}(\lambda_i)+\text{sgn}(\lambda_i^*)]$ which guarantees the decoupling of positive and negative semilattices. Furthermore, they also are the reason behind the superselection of the Hilbert subspaces with support on the eight octants defined by the signs of $\lambda_i$. The key point is that the action of $\hat{F}_i$ only produces changes in the sign of \emph{one} of the $\lambda$-variables. In the $v$-representation, this is translated into the realization that the action of $\hat{F}_i$ can only produce either a change in the sign of $v$ alone (that is, a change in the sign of $\lambda_\theta$) or a simultaneous change in the signs of $v$ and one of the anisotropy variables. In the case of the Lorentzian part, however, this is no longer true. Indeed, the argument goes as follows. The ${\Theta}$-operators (take $\hat{\Theta}_\theta$, for example) can produce changes in the three labels of the state they act upon (but never more than two at a time). This is a result of the appearance of holonomy elements corresponding to all spatial directions in Eq. \eqref{Thetai}. Nevertheless, the factor $[\text{sgn}(\lambda_i)+\text{sgn}(\lambda_i^*)]$ that weights each term in the action of $\hat{G}_i$ only protects the sign of one of them. The others could \emph{a priori} change sign without this factor noticing it. While all possible changes of sign were linked together in the case of the Euclidean part (a fact which leads to the protection of the sign of the three labels and the subsequent decoupling of the eight octants), the action of the Lorentzian part does not exhibit this property. Consider, for instance, the state corresponding to $l=m=-1$ that appears in the first line of Eq. \eqref{Thetatheta} provided that $v\neq2,4$,
\begin{align}
\left[1+\text{sgn}\left(v-2\right)\right]\bigg|{v-2-2\,\text{sgn}(v-2),\dfrac{v-2-2\,\text{sgn}(v-2)}{v-2}\lambda_\sigma,\lambda_\delta}\bigg\rangle,
\end{align}
where the associated weighting factor has been written down as well. If $2<v<4$, then, $v-2>0$ and the considered state has a nonvanishing contribution. However, the two first labels of the state $\ket{v,\lambda_\sigma,\lambda_\delta}\in\text{Cyl}_{\text{S}}^+$, affected by $\hat{G}_\theta$, have changed sign. Thus, at least one of the states connected with it through the action of $\hat{G}_\theta$ belongs to a Hilbert subspace with support on a different octant (namely, the one corresponding to $v<0$, $\lambda_\sigma<0$, and $\lambda_\delta>0$). In conclusion, the Hilbert subspaces with support on one of the eight octants defined by the signs of the triad variables are not left invariant by $\hat{\mathscr{H}}_L^{\text{BI}}$. In particular, the action of $\hat{\mathscr{H}}_L^{\text{BI}}$ connects the discrete semilattices of positive and negative $v$.

Before discussing the precise way in which the superselection sectors are enlarged with respect to the Euclidean case, we will comment on the action of $\hat{\Theta}_\sigma$ for completeness (the action of the remaining $\Theta$-operator is analogous to the one presented below). Particularizing for the case where $\ket{v,\lambda_\sigma,\lambda_\delta}\in\text{Cyl}_{\text{S}}^+$, the action of $\hat{\Theta}_\sigma$ is
\begin{align}
\hat{\Theta}_\sigma\ket{v,\lambda_\sigma,\lambda_\delta}=&-\dfrac{i}{8}\sum_{(l,m)\in I_v} \left\{l\,\bigg|{v+2l+2m\,\text{sgn}(v+2l),\dfrac{v+2l}{v}\lambda_\sigma,\lambda_\delta}\bigg\rangle\right.\nonumber\\
&+m\,\bigg|{v+2l+2m\,\text{sgn}(v+2l),\dfrac{v+2l+2m\,\text{sgn}(v+2l)}{v+2l}\lambda_\sigma,\lambda_\delta}\bigg\rangle\nonumber\\
&+l\,\bigg|{v+2l+2m\,\text{sgn}(v+2l),\dfrac{v+2l}{v}\lambda_\sigma,\dfrac{v+2l+2m\,\text{sgn}(v+2l)}{v+2l}\lambda_\delta}\bigg\rangle\nonumber\\
&\left.+m\,\bigg|{v+2l+2m\,\text{sgn}(v+2l),\dfrac{v+2l+2m\,\text{sgn}(v+2l)}{v+2l}\lambda_\sigma,\dfrac{v+2l}{v}\lambda_\delta}\bigg\rangle\right\}.\label{Thetasigma}
\end{align}

It is straightforward to realize that there are terms that may lie out of the Hilbert subspace of the original state. When we compute the action of $\hat{G}_\sigma$, each of the states displayed above comes preceded by a factor of the form $[1+\text{sgn}(\lambda_\sigma^*)]$ (where $\lambda_\sigma^*$ denotes the second label of each state). Therefore, changes in the sign of this anisotropy variable are prevented. Nevertheless, changes in the sign of $v$ or $\lambda_\delta$ could occur if they do not happen together with a change in the sign of $\lambda_\sigma$. It is clear that this situation does indeed take place (at least when $v\neq 2,4$). Consider, for instance, the target state corresponding to $l=m=-1$ in the first line of Eq. \eqref{Thetasigma}:
\begin{align}
[1+\text{sgn}(v-2)]\bigg|{v-2-2\,\text{sgn}(v-2),\dfrac{v-2}{v}\lambda_\sigma,\lambda_\delta}\bigg\rangle,
\end{align}
where we have included the factor that accompanies it within $\hat{G}_\sigma$, up to a multiplicative constant of $1/2$. If the original state is such that $2<v<4$, then $v-2>0$ and the above contribution does not vanish, because there is no change of sign in its label $\lambda_{\sigma}$. However, its volume $v^*=v-4$ is negative, so that this state no longer lives in the Hilbert subspace with support on the first octant.

We can also find instances where the change of sign also occurs in the label $\lambda_\delta$. For example, consider the state corresponding to $l=m=-1$ in the third line of Eq. \eqref{Thetasigma}:
\begin{align}
[1+\text{sgn}(v-2)]\bigg|{v-2-2\,\text{sgn}(v-2),\dfrac{v-2}{v}\lambda_\sigma,\dfrac{v-2-2\,\text{sgn}(v-2)}{v-2}\lambda_\delta}\bigg\rangle.
\end{align}
Once more, in the case where $2<v<4$, the prefactor of this contribution does not vanish, but both $v^*$ and $\lambda_\delta^*$ are negative.

\subsection{Superselection sectors}\label{sup}

Once we have completed the description of the $\Theta$-operators, from which the action of $\hat{G}_i$ can be immediately derived, we conclude this section with an analysis of the superselection sectors of the Lorentzian part of the Hamiltonian constraint operator. As we have proven above, the superselection sectors of the Euclidean part are not generically left invariant under the action of the Lorentzian Hamiltonian, which means that the superselection sectors of the Dapor-Liegener model of Bianchi I LQC are larger than the ones encountered in the standard formalism. In particular, we have noticed that the action of the Lorentzian Hamiltonian actually induces changes in the signs of the triad variables, which implies that the kinematical Hilbert space does not split into superselected Hilbert subspaces with support on each of the octants as in Refs. \cite{Bianchii,Bianchiii}. 

Nevertheless, the Lorentzian Hamiltonian has an action which is very similar to that of the Euclidean Hamiltonian, inasmuch as it produces constant shifts in the volume variable and dilations/contractions in the anisotropy variables. While the Euclidean part is characterized by producing shifts of (plus or minus) four units in $v$, if any, the Lorentzian part leads to shifts which are twice as large. The fact that the size of the shifts produced by $\hat{\mathscr{H}}_L^{\text{BI}}$ is an integer multiple of the size of the shifts produced by $\hat{\mathscr{H}}_E^{\text{BI}}$ ensures that, as far as the volume is concerned, the kinematical Hilbert space is still superselected in invariant Hilbert subspaces with support on discrete lattices of step four. Even though the positive and negative semilattices are not decoupled, a whole discrete lattice can always be written as a direct sum of two semilattices. Namely, if the initial state has a volume $v=\varepsilon+4n$, for some $n\in\mathbb{N}$, then it is immediate to show that the repeated action of $\hat{\mathscr{H}}_L^{\text{BI}}$ (and of the whole Hamiltonian constraint, for that matter) connects this state with states $\ket{v^*,\lambda_\sigma^*,\lambda_\delta^*}$ such that $v^*$ belongs to $\mathcal{L}_{\varepsilon}^+\cup \mathcal{L}_{4-\varepsilon}^-$, where $\mathcal{L}_{\tilde{\varepsilon}}^-=\{v=-(\tilde{\varepsilon}+4n):\, n\in\mathbb{N}\}$. The only exceptions are found for $\varepsilon=2,4$. In these special cases, the projections onto the Hilbert completion of $\widetilde{\text{Cyl}}_{\text{S}}$ ensure that the states of vanishing volume are totally decoupled, even at intermediate stages (i.e. after the action of a single holonomy). This implies that negative volumes can never be reached from $v=2$ or $4$. Indeed, for this to happen, it would be necessary to pass over a singular state, something that is precluded by the projections $\hat{\mathcal{P}}$ or, equivalently, by the restriction $(l,m)\in I_v$. Thus, in these particular cases, the volume lies on discrete \emph{semi}lattices, $\mathcal{L}_2^+$ or $\mathcal{L}_4^+$, instead.

A similar situation is encountered in the anisotropy sector. In the Euclidean case, in each section of constant volume, only a subspace of states are connected with $\ket{v,\lambda_\sigma,\lambda_\delta}$ by the repeated action of $\hat{\mathscr{H}}_E^{\text{BI}}$, namely the ones with quantum numbers $\lambda_\sigma^{*}$ and $\lambda_\delta^{*}$ that are of the form $\lambda_r^{*}=w_{\varepsilon}\lambda_r$, where $w_\varepsilon\in\mathcal{W}_{\varepsilon}$. This can no longer be true when the Lorentzian part of the Hamiltonian constraint is regularized independently. Indeed, this is due to the fact that the anisotropy variables can change sign under the action of $\hat{\mathscr{H}}_L^{\text{BI}}$. This change, however, does not alter completely the structure of the superselection sectors. The reason behind this fact can be understood straightforwardly. Even if the functional form of the Lorentzian Hamiltonian is considerably more involved than that of the Euclidean one, they only differ in that a larger number of holonomy elements $\hat{\mathcal{N}}_{\pm2\bar{\mu}_i}$ act on each state in the Lorentzian case. However, these holonomy elements act in a very simple manner: they produce shifts of two units in $v$ and, when $i=\sigma$ or $\delta$, they also produce dilations/contractions in the anisotropy variables of the form $\lambda_r\rightarrow (v^*/v)\lambda_r$. Thus, the anisotropy variables corresponding to states connected with $\ket{v,\lambda_\sigma,\lambda_\delta}$ via the action of the Hamiltonian can be obtained from the original anisotropies through the multiplication of a number of fractional factors, in which the numerator and denominator differ by two units. In this sense, as in the Euclidean case, the effect caused on the anisotropies only depends on $v$ and, therefore, ultimately only on $\varepsilon$ [see e.g. Eqs. \eqref{Thetatheta} and \eqref{Thetasigma}]. However, in the same way that the volume can now change sign, a parallel situation is found for the anisotropy variables. The restriction on $z$ in the definition of $\mathcal{W}_{\varepsilon}$ [given by Eq. \eqref{W+}] that ensures that the fractional factors are positive disappears when the Lorentzian part is included in the Hamiltonian. In this new situation, the coefficient $\tilde{w}_\varepsilon$ that characterizes the action on the anisotropies belongs to a larger set $\tilde{\mathcal{W}}_\varepsilon$, defined by
\begin{align}
\tilde{\mathcal{W}}_\varepsilon=\left\{\prod_{n,m\in\mathbb{Z}}\left(\dfrac{\varepsilon+2n}{\varepsilon+2m}\right)^{k^n_m}:\, k^n_m\in\mathbb{N};\ n,m\neq-\varepsilon/2\, \text{ if } \varepsilon=2,4\right\},
\end{align}
where the last requirement avoids singular fractional factors in the product, something that is guaranteed by the functional form of the Hamiltonian constraint, as we have have already discussed.

The elements of this set turn out to admit a simple expression in terms of those of $\mathcal{W}_{\varepsilon}$. The fractional factors can be rearranged in such a way that the elements of $\tilde{\mathcal{W}}_\varepsilon$ can be written as
\begin{align}\label{wtilde}
\tilde{w}_\varepsilon=C_{\varepsilon}w_{\varepsilon}w_{\varepsilon^{\prime}},
\end{align}
where $C_{\varepsilon}$ is a coefficient that dictates the global sign of $\tilde{w}_\varepsilon$. Moreover, $w_{\varepsilon}$ and $w_{\varepsilon^{\prime}}$ belong to $\mathcal{W}_\varepsilon$ and $\mathcal{W}_{\varepsilon^{\prime}}$, respectively, where we have introduced the notation
\begin{equation}
\varepsilon^{\prime}=\left\{   
\begin{array}{l}
4-\varepsilon \quad \; \forall\,\varepsilon \neq 2, 4, \\
\varepsilon \quad \qquad {\rm if} \; \varepsilon =2, 4.
\end{array}
\right.
\end{equation}
The origin of the coefficient $C_{\varepsilon}$ lies on the possible existence of (integer powers of) a fractional factor with numerator and denominator of different signs. Explicitly,
\begin{align}\label{C}
C_{\varepsilon}=\left\{\begin{array}{cl}
(-1)^{z'}\bigg|\dfrac{\varepsilon-2}{\varepsilon-2-2\,\text{sgn}(\varepsilon-2)}\bigg|^{z'}&\text{with}\ z'\in\mathbb{Z}, \; \varepsilon \neq 2, 4,\\
1&\text{if}\ \varepsilon=2,4.
\end{array}
\right.
\end{align}

The peculiarity of the cases $\varepsilon=2,4$ is again due to the decoupling of the states of vanishing volume in the Lorentzian part of the Hamiltonian constraint. This decoupling implies that there can exist no fractional factor with numerator and denominator of different signs, since positive and negative volumes cannot be connected when $\varepsilon=2$ or $4$. As a result, the sign of the anisotropies is respected as well. Therefore, the superselection sectors of the Euclidean Hamiltonian are not enlarged in these cases. It is simple to realize that this conclusion is already manifest in our definitions. Indeed, in the special cases $\varepsilon=2,4$ we see that $\mathcal{W}_\varepsilon=\mathcal{W}_{\varepsilon^{\prime}}=\mathcal{W}_\varepsilon\cdot \mathcal{W}_{\varepsilon^{\prime}}$, the last equality being a direct consequence of the definition of $\mathcal{W}_\varepsilon$, that guarantees that any product of its elements still  belongs to that set. Using this and the fact that $C_{\varepsilon}=1$, one can straightforwardly see that $\tilde{\mathcal{W}}_\varepsilon=\mathcal{W}_\varepsilon$ in these specific cases.

Finally, it is worth remarking some properties of $\tilde{\mathcal{W}}_\varepsilon$. Although this set is generically larger than its Euclidean counterpart $\mathcal{W}_{\varepsilon}$, it happens to be still \emph{countably} infinite (as it is clear from its definition) and dense, this time in the whole real line. One can prove the latter of these statements in a straightforward way by following a procedure entirely analogous to the one presented in Appendix D of Ref. \cite{Gowdy1}. Indeed, the subset of positive $\tilde{w}_{\varepsilon}$ is trivially dense on $\mathbb{R}^{+}$, since it contains $\mathcal{W}_\varepsilon$, which is dense \cite{Gowdy1}. In a similar manner, it is simple to show that the same holds for the subset of negative $\tilde{w}_{\varepsilon}$ on $\mathbb{R}^{-}$.

In summary, even though the independent regularization of the Lorentzian part of the Hamiltonian constraint introduces difficulties in the formalism of anisotropic LQC, the superselection sectors of the full Hamiltonian partially retain the structure of those of the Euclidean part alone. Indeed, they can be decomposed in terms of the simpler superselection sectors encountered in the standard approach to Bianchi I LQC. We have reasoned that the quantum number $v$ generically belongs to a discrete lattice of step four, which can be viewed as the direct sum of its positive and negative semilattices. Moreover, in each section of constant $v$, only certain states are connected via the repeated action of the Lorentzian Hamiltonian. Such states are characterized by anisotropy variables that are proportional to the original ones, with a proportionality factor that only depends on $v$ and belongs to a set $\tilde{\mathcal{W}}_\varepsilon$. The elements of this set can again be written in terms of those of two similar sets that appeared in the definition of the superselection sectors of the Euclidean Hamiltonian. As mentioned above, even though the discrete lattices and $\tilde{\mathcal{W}}_{\varepsilon}$ are infinite sets, they are \emph{countable}. Hence, the action of the full Hamiltonian operator still superselects separable Hilbert subspaces.

\subsection{A comment on the isotropic scenario}\label{iso}

We have seen how, in the Dapor-Liegener formalism of Bianchi I cosmologies, the superselection sectors get enlarged with respect to those found in the standard formalism. However, this result might appear to be in tension with the conclusions about flat FLRW cosmologies discussed in Ref. \cite{DaPorLie}. In that isotropic case, we found that the superselection sectors of the standard formalism are indeed preserved under the modification of the Hamiltonian obtained with an independent regularization of the Lorentzian term (see Ref. \cite{DaPorLie} and the Appendix for further details). Hence, there must exist a mechanism that prevents the enlargement of the considered sectors in the isotropic scenario. 

As we have explained above, the superselection sectors change owing to the fact that the action of the Lorentzian part of the Hamiltonian does not leave invariant the sign of the components of the densitized triad. In other words, this Lorentzian part provides contributions with  quantum numbers that can have the opposite sign of those of the original state on which the Hamiltonian acts. In the Euclidean case, these flips of sign were prevented by a factor of the form $[\text{sgn}(\lambda_i)+\text{sgn}(\lambda_i^*)]$ that accompanies every single contribution in the action of $\hat{F}_i$. Since the $F$-operators only affect their respective $\lambda_i$ and leave the other two variables unchanged, the aforementioned factor suffices to ensure that any contribution that could potentially belong to another octant actually vanishes. Nevertheless, the analogs of these operators in the Lorentzian case, $\hat{G}_i$, do not display the same behavior. Indeed, since each of them contains products of holonomy elements along two different spatial directions, the $G$-operators can produce changes in two out of the three $\lambda$-variables. Given that only one sign is protected (the sign of $\lambda_i$ under the action of $\hat{G}_i$), the enlargement of the superselection sectors occurs when a contribution is such that $\lambda_i$ maintains its sign while the other affected label, $\lambda_{j}$ ($j\neq i$), experiences a flip of sign. Then, the mechanism that guarantees the decoupling of octants in the Euclidean case fails to be effective.

An interesting question is how this situation is resolved in the isotropic limit. In view of our previous comments, the answer is straightforward. In isotropic situations, all the nontrivial components of the densitized triad, on the one hand, and of the connection, on the other hand, coincide, leading to only one pair of canonical variables: $(c^i,p_i)\rightarrow (c,p)$. As a result, every $\lambda$-variable reduces to the same function of the isotropic phase space, $\lambda_i\rightarrow \lambda=\text{sgn}(p)|p|^{1/2}/(4\pi G\gamma\sqrt{\Delta})^{1/3}$. Therefore, the sign of any odd power of the $\lambda_i$'s reduces under isotropy to the sign of $p$ or, equivalently, the sign of $v$. Then, in this situation, every contribution in the action of the Lorentzian part of the Hamiltonian appears weighted by a factor $[\text{sgn}(v)+\text{sgn}(v^*)]$. This factor vanishes identically when $v$ (which is the only variable that labels states in flat FLRW cosmologies) changes sign. Hence, no contribution can lead to a state with an opposite orientation in this isotropic regime, and positive and negative volumes decouple under the action of the Hamiltonian. This is the reason behind the persistence of the decoupling mechanism in the isotropic case: there is a single triad variable (recall that the three spatial directions are identified in this scenario) and, therefore, there is no extra triadic degree of freedom that may flip its sign without $v$ noticing it.

\section{Conclusions and discussion}\label{discussion}

Recently, considerable attention has been paid to studying the mathematical ambiguities that affect the formulation of LQC and, in particular, to explore alternatives to the standard regularization of the Hamiltonian constraint \cite{TT,ALM,AAL}. One of these alternatives has reached a more prominent status, because it is especially interesting from the physical point of view. Some authors refer to it as the Dapor-Liegener formalism of LQC \cite{DL1,DL2} and it is based on a procedure to regularize the Hamiltonian constraint which is conceptually identical to the one used in LQG. Indeed, in this formalism the Euclidean and Lorentzian parts of the Hamiltonian constraint are regularized independently. This is in contrast with the procedure used in the standard formalism of LQC \cite{APS2,MMO}, which appeals to symmetry considerations in order to reexpress the Lorentzian part in terms of the Euclidean one before regularization and, therefore, cannot be applied to more general scenarios without a substantial modification. This conceptual drawback of the standard formalism, together with the fact that we do not yet fully understand the relation between LQG and LQC \cite{Engle,BK,Engle2,Paw,BEHM}, makes the Dapor-Liegener proposal particularly attractive and, during the past year, efforts have been made to extract the physical implications entailed by such a modification of the Hamiltonian constraint. Although many aspects have been analyzed, the studies carried out so far have focused on the case of homogeneous and isotropic cosmologies \cite{YDM,DL1,DL2,DL3,Paramc1,Paramc2,genericness,Agullo,gaugeinvariant,CSaction,Haro,DaPorLie}. 

The aim of this article is to extend the scope of the already existing analyses by including anisotropies in the cosmological spacetimes under consideration. The simplest anisotropic spacetimes, which exhibit spatial homogeneity, are the Bianchi I cosmologies \cite{kramer}. These cosmologies have been thoroughly studied by applying polymeric quantization techniques \cite{chiou1,chiou2,AW-E1,Bianchii,Bianchiii}. However, the quantization of Bianchi I cosmologies within the framework of LQC has been implemented so far using exclusively the standard regularization procedure. In other words, spatial flatness and homogeneity have been used to symmetry-reduce the Hamiltonian before its regularization, leading to an expression solely in terms of the Euclidean part. Although we already discussed the regularization of the Lorentzian part in Ref. \cite{DaPorLie}, no detailed study had been carried out until present at the quantum level. This article introduces, for the first time in the literature, an independent treatment  of the Lorentzian part of the Hamiltonian constraint of Bianchi I spacetimes, including its quantum representation and the analysis of some of the properties of the quantum theory constructed in this way.

After reviewing the quantum kinematics of the Bianchi I loop quantum spacetimes in Sec. \ref{QK}, we have briefly revisited the quantization of the Euclidean part of the Hamiltonian constraint using the MMO quantization prescription \cite{MMO} (see Sec. \ref{QEH}). More concretely, we have emphasized how the kinematical Hilbert space splits into different sectors under the action of the Euclidean Hamiltonian. This occurs because the signs of the quantum numbers that label the states are conserved. In turn, this is an immediate consequence of a defining feature of the MMO prescription: the signs of the triad variables and the holonomies are symmetrized upon quantization. As a result of this prescription, the kinematical Hilbert space is superselected into eight Hilbert subspaces. Furthermore, owing to the precise functional form of the Euclidean part, its action only produces discrete shifts in the volume, while the anisotropies undergo dilations and contractions that only depend on this volume. Thus, the volume variable lives on a discrete semilattice and, in each section of constant volume, the states that are connected by the (repeated) action of the Euclidean Hamiltonian have anisotropies that can be characterized by a proportionality factor which necessarily belongs to a certain set. This set is considerably more involved than the semilattices appearing in the volume sector. The resulting picture is that the Euclidean part of the Hamiltonian (and, thus, the standard Hamiltonian in LQC) obtained via the implementation of the MMO prescription has a well-defined quantum representation and its action leaves invariant a number of superselection sectors that are well understood. However, the complexity of its functional form has frustrated any attempt at determining its spectral properties, at least in the case that has been presented in this article. It is important to emphasize that there exist other ambiguities in the quantization program, like e.g. those related with a specific choice for the implementation of the improved dynamics \cite{APS2}. As we commented in Sec. \ref{QK}, there exists consensus in the community concerning the improved dynamics in isotropic scenarios, although several proposals have been suggested in principle for its implementation in anisotropic scenarios. Among them, nevertheless, there is one which is broadly accepted \cite{AW-E1}, and it is this proposal that we have employed throughout this work. Nonetheless, it is worth commenting that this is not the only existing option \cite{chiou1}. We will discuss some of the consequences of our choice at the end of this section, after summarizing the main results of this paper.

We have also investigated whether the most important features of the LQC description of Bianchi I spacetimes obtained with the standard formalism remain present when the Lorentzian part of the Hamiltonian is regularized and quantized independently. To this effect, in Sec. \ref{QLH} we have begun by promoting the classical Lorentzian Hamiltonian to an operator acting on the kinematical Hilbert space, with a rather general choice of algebraic factor ordering for the volume operator. As in the Euclidean case, the quantum analogs of the classical singularities are found to decouple. After a densitization of the Lorentzian contribution, we have proceeded to the explicit computation of the action of the basic operators that appear in the resulting expression. We have found that the adopted factor ordering for the volume operator ensures that singular states do not contribute to the action of the Lorentzian Hamiltonian. This fact is rooted in a number of effective projections onto the orthogonal complement of the kernel of the volume, operative even at intermediate stages. Then, we have motivated the choice of a particular factor ordering for the volume that is compatible with the prescriptions put forward in previous works on the Dapor-Liegener formalism.  With this ordering, we have investigated the subsequent action of the operators that form the densitized Lorentzian Hamiltonian with the objective of extracting information about the decoupling of the octants and the structure of the superselection sectors. Whereas the action of these operators has been found to be similar to that of their Euclidean counterparts, an essential difference has been noted as far as the volume is concerned: they result in shifts which are twice as large as those produced by the Euclidean operators. Consequently, the whole Lorentzian part produces shifts of eight units (if any) in the volume, as opposed to the four-unit shifts generated by the Euclidean part.

Once these properties of the action of the Lorentzian Hamiltonian have been established, we have discussed the corresponding superselection rules in Sec. \ref{sup}. Since the shifts in the volume are twice larger, one may think that the action of the Lorentzian part leaves invariant the discrete semilattices that were superselected by the Euclidean part. However, this is not the case. The reason behind this crucial distinction is that the operators mentioned above do not preserve the sign of the triad variables, a fact which precludes the decoupling of octants. In particular, the volume can change sign under the action of the Lorentzian Hamiltonian. As a result, \emph{whole} discrete lattices of step four (which can always be written as the direct sum of two semilattices: a positive and a negative one) are now superselected. Something similar occurs with the anisotropy labels. Indeed, the repeated action of the Lorentzian part alters the anisotropy variables by dilating and contracting them via a factor which is given by the product of a number of fractions, with numerators and denominators that differ by two units. Such fractional factors only depend on the volume. Up to this point, our result is identical to the one obtained when only the Euclidean part is considered. However, just as the volume can change sign under the action of the Lorentzian part, so can the anisotropy variables. Hence, the set involved in the definition of the superselection sectors for the anisotropies must be enlarged to encompass negative values, although the underlying structure remains unchanged with respect to the one appearing in the Euclidean case. 

We have also discussed the exceptional cases of the lattices that are symmetric with respect to the origin, which we had deliberately left out in the previous paragraph. Because of the projections onto the completion of $\widetilde{\text{Cyl}}_{\text{S}}$ that, in practice, operate after the action of each holonomy element, we have found that the positive and negative semilattices actually decouple. Thus, in those special cases, the superselection sectors of the standard formalism are preserved under the independent regularization of the Lorentzian term.

The introduction of the Lorentzian part complicates substantially the functional form of the Hamiltonian constraint operator. Although the large number of terms that determine its action on the elements of the volume eigenbasis makes impractical to write down this action explicitly, we have been able to extract information about the structure of this constraint operator and about its superselection sectors. In fact, its behavior turns out to be similar to that of the Euclidean constraint. As we have commented, the two key differences are that: i) the Lorentzian part produces shifts in the volume which are twice larger, and ii) the signs of the triad variables are not respected. The latter leads to a generic enlargement of the superselection sectors, which seems to indicate that the inclusion of the Lorentzian term in the regularization procedure cannot be regarded as a mere higher-order correction. 

To conclude, we would like to add a comment on the issue of the implementation of the improved dynamics prescription. In this respect, it seems worthy to consider the consequences of adopting an alternative prescription like, e.g., that of Ref. \cite{chiou1}. In standard LQC, where the Hamiltonian is expressed as purely Euclidean, such a prescription results in a quantum theory which is considerably simpler, compared to that of Ref. \cite{AW-E1}. In the following, we will refer to these two different proposals as ``scheme A'' and ``scheme B'', respectively. In scheme A, the minimum coordinate length in each spatial direction only depends on the triad variable corresponding to the same spatial direction. This is in contrast with the case of scheme B, where each minimum coordinate length depends on all triad variables. This difference leads to a number of simplifications. For instance, the holonomy elements associated with a certain spatial direction only produce shifts in the same spatial direction, a behavior that makes the quantum theory separable in three identical copies, one per direction. A related property is that the resulting terms in the Euclidean Hamiltonian corresponding to each of the three directions commute, a fact that implies that each of them is conserved as a Dirac observable. 

At the kinematical level, scheme A allows us to define volume-like variables for all spatial directions, instead of a single volume variable and two anisotropy variables. In such a representation, all holonomy elements produce constant shifts in their corresponding volume variable. Furthermore, since holonomy elements along different spatial directions commute, so do the projected sine and cosine operators corresponding to different spatial directions. This simplifies considerably the action of the $G$-operators, reducing the number of terms in their action by half. Nevertheless, even when this simpler quantization prescription is adopted, Hilbert subspaces with support on different octants generically fail to be decoupled. The reason why is obvious. Indeed, only the sign of a single triad variable is protected by the symmetrization of the sign and the holonomy elements in each term of the Hamiltonian. However, because of the simultaneous appearance of holonomy elements corresponding to two different spatial directions per term, sign flips can be induced in a triad variable for which the sign is not protected. Therefore, the action of the Lorentzian part does not decouple the Hilbert subspaces with support on different octants and the superselection sectors get enlarged with respect to those of the standard regularization, like in the case of the improved dynamics prescription considered in most of our analysis.

\appendix
\section{Alternative orderings in the isotropic theory}\label{app} 

In this Appendix, we present a reformulation of the Dapor-Liegener model of flat FLRW LQC, originally quantized adopting the MMO prescription in Ref. \cite{DaPorLie}. The factor ordering of this reformulation  makes it directly comparable with the analysis developed in the present article. We will discuss two types of alternative quantization choices with respect to Ref. \cite{DaPorLie}: the definition employed to represent $1/\bar{\mu}$ and the adoption of a general algebraic factor ordering for the powers of the volume operator.

In the following, we consider the isotropic case, so that all three spatial directions are identified and, therefore, all three canonical pairs $(c^i,p_i)$ reduce to a single pair $(c,p)$, that describes the isotropic gravitational degrees of freedom, supplied with the nontrivial Poisson bracket $\{c,p\}=8\pi G\gamma/ 3$. Quantum states are labeled by a single quantum number $v$ and the action of the fundamental operators on the orthonormal $v$-basis of $\mathcal{H}_{\text{iso}}^{\text{kin}}$ is
\begin{equation}
\hat{V}\ket{v}=2\pi G\gamma\sqrt{\Delta}|v|\ket{v},\quad
\hat{\mathcal{N}}_{\pm n\bar{\mu}}\ket{v}=\ket{v\pm n},
\end{equation}
where $\bar{\mu}$ is the minimum coordinate length, given by $\bar{\mu}=\sqrt{\Delta /|p|}$.

In this kind of systems, the classical gravitational contributions to the  Hamiltonian constraint can be obtained from Eqs. \eqref{HEreg} and \eqref{HLreg} by taking the isotropic limit (see Ref. \cite{DaPorLie} for further details). The resulting expressions are
\begin{align}
H_E^{\text{FLRW}}&=\dfrac{3}{8\pi G V}\left(\text{sgn}(p)|p|\dfrac{1}{\bar{\mu}}\sin\bar{\mu}c\right)^2,\\
H_L^{\text{FLRW}}&=-\dfrac{3}{8\pi GV}\dfrac{1+\gamma^2}{4\gamma^2}\left(2\,\text{sgn}(p)|p|\dfrac{1}{\bar{\mu}}\sin\bar{\mu}c\,\cos\bar{\mu}c\right)^2.
\end{align}
According to the MMO prescription, we represent the Euclidean and Lorentzian contributions by operators on $\mathcal{H}^{\text{kin}}_{\text{iso}}$ of the form
\begin{align}
\hat{H}_{E,L}^{\text{FLRW}}=\left[\widehat{\dfrac{1}{V}}\right]^{1/2}\hat{\mathscr{H}}_{E,L}^{\text{FLRW}}\left[\widehat{\dfrac{1}{V}}\right]^{1/2}.
\end{align}
As in Secs. \ref{QEH} and \ref{QLH}, we perform a change of densitization and restrict our analysis to the study of the densitized operators defined on the orthogonal complement of the singular state, $\widetilde{\mathcal{H}}^{\text{kin}}_{\text{iso}}$. In Ref. \cite{DaPorLie}, the densitized operators were given by
\begin{equation}\label{HdensFLRW}
\hat{\mathscr{H}}_E^{\text{FLRW}}=\dfrac{3}{8\pi G}\hat{\Omega}_{2\bar{\mu}}^2,\quad
\hat{\mathscr{H}}_L^{\text{FLRW}}=-\dfrac{3}{8\pi G}\dfrac{1+\gamma^2}{4\gamma^2}\hat{\Omega}_{4\bar{\mu}}^2,
\end{equation}
where, for any integer value of $n$,
\begin{align}\label{Omegan}
\hat{\Omega}_{n\bar{\mu}}=\dfrac{1}{4i\sqrt{\Delta}}\left[\widehat{\dfrac{1}{\sqrt{|p|}}}\right]^{-1/2}\widehat{\sqrt{|p|}}\left[\widehat{\text{sgn}(p)},\hat{\mathcal{N}}_{n\bar{\mu}}-\hat{\mathcal{N}}_{-n\bar{\mu}}\right]_{+}\widehat{\sqrt{|p|}}\left[\widehat{\dfrac{1}{\sqrt{|p|}}}\right]^{-1/2}.
\end{align}

The main difference between Ref. \cite{DaPorLie} and the quantization adopted here resides in how $1/\bar{\mu}$ is represented as an operator within the Hamiltonian constraint. In Ref. \cite{DaPorLie}, we employed the definition
\begin{align}
\widehat{\dfrac{1}{\bar{\mu}}}=\dfrac{1}{\sqrt{\Delta}}\left[\widehat{\dfrac{1}{\sqrt{|p|}}}\right]^{-1}.
\end{align}
This is the most commonly used definition in isotropic LQC (see e.g. Refs. \cite{APS2,MMO}). In anisotropic cosmologies (and, in particular, in Bianchi I) the most familiar convention fails to recover exactly the above definition in the isotropic scenario. The isotropic counterpart of the definition adopted in this paper (and in other analyses of  Bianchi I spacetimes in the literature, like e.g. Refs. \cite{Bianchii,Bianchiii,AW-E1}) is
\begin{align}\label{defmu}
\widehat{\dfrac{1}{\bar{\mu}}}=\dfrac{1}{\sqrt{\Delta}}\widehat{\sqrt{|p|}}.
\end{align}
This latter choice of representation leads to the appearance of powers of the volume in the densitized Hamiltonian [see, for instance, Eqs. \eqref{HEidens} and \eqref{HLop}], whereas in Ref. \cite{DaPorLie} powers of $\widehat{\sqrt{|p|}}$ and $\widehat{1/\sqrt{|p|}}$ appeared instead [see Eq. \eqref{Omegan}]. In this Appendix, we will rewrite the full quantum Hamiltonian constraint for flat FLRW cosmologies using Eq. \eqref{defmu} and show that the main results of Ref. \cite{DaPorLie} are robust against this change. Furthermore, we will also consider a general factor ordering for the powers of the volume operator of the kind discussed in the beginning of Sec. \ref{QLH}.

The densitized versions of the Euclidean and Lorentzian parts of the Hamiltonian have an expression similar to Eq. \eqref{HdensFLRW} except for the fact that the functional form of $\hat{\Omega}_{n\bar{\mu}}$ is altered, owing to the modifications in our representation. We denote the new operators by $\hat{\tilde{\Omega}}_{n\bar{\mu}}^{(\alpha)}$, with $n=2,4$, where $\alpha$ is again the parameter that labels the choice of factor ordering for the powers of the volume. Let us write those operators down explicitly and study their action on $\widetilde{\mathcal{H}}^{\text{kin}}_{\text{iso}}$. 

In the case of the Euclidean part, namely $n=2$, only one holonomy operator is involved in each term of $\hat{\tilde{\Omega}}_{2\bar{\mu}}^{(\alpha)}$, and therefore it seems natural to order the powers of the volume in an algebraically symmetric way to the left and right of the holonomy. In this manner, we arrive at a specific factor ordering and the ambiguity parametrized by $\alpha$ does not appear. In our notation, this is equivalent to setting $\alpha=0$ in the Euclidean part. On the other hand, using the definition \eqref{defmu}, $|p|/\bar{\mu}$ is represented by $\hat{V}/\sqrt{\Delta}$, so that
\begin{align}\label{Omeganew}
\text{sgn}(v)|p|\dfrac{1}{\bar{\mu}}\sin\bar{\mu}c\to \hat{\tilde{\Omega}}_{2\bar{\mu}}^{(0)}=\dfrac{1}{2\sqrt{\Delta}}\left[\widehat{\text{sgn}(v)},\hat{V}^{1/2}\widehat{\sin\bar{\mu}c}\,\hat{V}^{1/2}\right]_{+}.
\end{align}
Introducing the notation
\begin{align}
\tilde{f}^{(n,\alpha)}_{\pm}(v)&=\dfrac{\pi G\gamma}{2}s_{\pm}^{(n)}(v)|v|^{1/2-\alpha}\bigg|v\pm \dfrac{n}{2}\bigg|^{2\alpha}|v\pm n|^{1/2-\alpha},\\
s_{\pm}^{(n)}(v)&=\text{sgn}(v)+\text{sgn}(v\pm n),
\end{align}
the action of the operator \eqref{Omeganew} on a volume eigenstate turns out to be
\begin{align}
\hat{\tilde{\Omega}}_{2\bar{\mu}}^{(0)}\ket{v}=-i\left\{\tilde{f}_+^{(2,0)}(v)\ket{v+2}-\tilde{f}_-^{(2,0)}(v)\ket{v-2}\right\}.
\end{align}
It is then immediate to see that the action of the densitized Euclidean contribution on $\widetilde{\mathcal{H}}^{\text{kin}}_{\text{iso}}$ is 
\begin{align}\label{ActionHEtilde}
\dfrac{8\pi G}{3}\hat{\tilde{\mathscr{H}}}_E^{\text{FLRW}}\ket{v}=-\tilde{f}^{(2,0)}_{+}(v)\tilde{f}^{(2,0)}_{+}(v+2)\ket{v+4}+\left\{\left[\tilde{f}^{(2,0)}_{+}(v)\right]^2+\left[\tilde{f}^{(2,0)}_{-}(v)\right]^2\right\}\ket{v}-\tilde{f}^{(2,0)}_{-}(v)\tilde{f}^{(2,0)}_{-}(v-2)\ket{v-4}.
\end{align}
This action has the same structure that one obtains using the other definition of the inverse of $\bar{\mu}$ (see Refs. \cite{MMO,DaPorLie}): it is an equation in finite differences which relates a volume eigenstate with two other eigenstates via constant shifts of four units. Consequently, Hilbert spaces with support on discrete lattices of step four are left invariant under the action of the Euclidean Hamiltonian.
 
 The functions $\tilde{f}$ that weight each of the possible contributions display a series of properties that are shared by their analogs in Refs. \cite{MMO,DaPorLie}. Indeed, these functions satisfy 
 \begin{equation}
 \tilde{f}^{(n,\alpha)}_{-}(v)=0\quad\forall\,v\in(0,n],\quad
 \tilde{f}^{(n,\alpha)}_{+}(v)=0\quad\forall\,v\in[-n,0),
 \end{equation}
owing to the presence of the linear combinations of signs $s^{(n)}_{\pm}(v)$, which are characteristic of the MMO prescription. As a result, the coefficient that weights an eigenvolume contribution vanishes if such a contribution experiences a flip of sign in $v$. This phenomenon decouples the sectors of positive and negative volumes and, thus, the action of the Euclidean part of the Hamiltonian constraint superselects separable Hilbert subspaces with support on discrete semilattices of step four, obtained by the Cauchy completion of  $\text{span}\{\ket{v}:\,v\in\mathcal{L}_\varepsilon^{\pm}\}$. 
 
Let us now study the Lorentzian contribution. Given that two holonomy operators appear in each term of this contribution, different factor orderings for the powers of the volume operator are possible, inserting part of them between the two holonomies that are present. We introduce a (potentially nonzero) parameter $\alpha$ to account for these choices. Following the rules of the MMO prescription, defining $1/\bar{\mu}$ as in Eq. \eqref{defmu}, and adopting a factor ordering for the volume analogous to the one employed in the beginning of Sec. \ref{QLH}, we get
\begin{align}
2\,\text{sgn}(v)|p|\dfrac{1}{\bar{\mu}}\sin\bar{\mu}c\,\cos\bar{\mu}c\to \hat{\tilde{\Omega}}_{4\bar{\mu}}^{(\alpha)}=\dfrac{1}{\sqrt{\Delta}}\left[\widehat{\text{sgn}(v)},\hat{V}^{1/2-\alpha}\,\widehat{\sin\bar{\mu}c}\,\hat{V}^{2\alpha}\,\widehat{\cos\bar{\mu}c}\,\hat{V}^{1/2-\alpha}\right]_+.
\end{align}
As in the Euclidean case, the limit $\alpha\to 0$ reproduces the construction of Ref. \cite{DaPorLie}, up to the changes induced by our different definition of $1/\bar{\mu}$. Nevertheless, we will mantain in our analysis a general value of $\alpha$, between 0 and 1/2, and particularize to the limit $\alpha\to 0$ only at the end.

It is straightforward to show that
\begin{align}
\hat{\tilde{\Omega}}_{4\bar{\mu}}^{(\alpha)}\ket{v}=-i\left\{\tilde{f}^{(4,\alpha)}_{+}(v)\ket{v+4}-\tilde{f}^{(4,\alpha)}_{-}(v)\ket{v-4}\right\}-i\tilde{g}^{(\alpha)}_{0}(v)\ket{v},
\end{align}
where
\begin{align}
\tilde{g}^{(\alpha)}_0(v)=-\pi G\gamma\,|v|^{1-2\alpha}\bigg||v-2|^{2\alpha}-|v+2|^{2\alpha}\bigg|.
\end{align}
As we can see, the action of the Lorentzian operator contains a term that produces no shift in the volume and that did not appear in previous considerations. However, it is worth noticing that the factor that weights this contribution vanishes in the limit $\alpha\to 0$. Then, in that case, we recover an action for our operator that is similar to the one obtained in Ref. \cite{DaPorLie} (except for the redefinition $f\rightarrow \tilde{f}$ that arises from the different representation of $1/\bar{\mu}$). 

The resulting action of $\hat{\tilde{\mathscr{H}}}_L^{\text{FLRW}}$ (where we omit the label $\alpha$ for the sake of brevity) is
\begin{align}\label{ActionHLtilde}
-\dfrac{8\pi G}{3}\dfrac{4\gamma^2}{1+\gamma^2}\hat{\tilde{\mathscr{H}}}_L^{\text{FLRW}}\ket{v}=&-\tilde{f}^{(4,\alpha)}_{+}(v)\tilde{f}^{(4,\alpha)}_{+}(v+4)\ket{v+8}+\left\{\left[\tilde{f}^{(4,\alpha)}_{+}(v)\right]^2+\left[\tilde{f}^{(4,\alpha)}_{-}(v)\right]^2\right\}\ket{v}\nonumber\\&-\tilde{f}^{(4,\alpha)}_{-}(v)\tilde{f}^{(4,\alpha)}_{-}(v-4)\ket{v-8}-i\tilde{g}_0^{(\alpha)}(v)\,\hat{\tilde{\Omega}}_{4\bar{\mu}}^{(\alpha)}\ket{v}\nonumber\\
&-\tilde{f}^{(4,\alpha)}_+(v)\tilde{g}_0^{(\alpha)}(v+4)\ket{v+4}+\tilde{f}^{(4,\alpha)}_-(v)\tilde{g}_0^{(\alpha)}(v-4)\ket{v-4}.
\end{align}
This action yields an equation in finite differences that relates a given volume eigenstate with a total of four other states, connected with it via constant shifts of four or eight units. In contrast with the situation found in Ref. \cite{DaPorLie}, there are now  terms with a shift of (plus or minus) four units in $v$ generated by the Lorentzian contribution. The origin of this discrepancy is the different factor ordering chosen for the powers of the volume. Indeed, note that the vanishing of $\lim_{\alpha\to 0} \tilde{g}_0^{(\alpha)}(v)$ makes all but the first three contributions disappear on the right hand side of the above expression, in the limit where we recover the usual factor ordering for the powers of the volume. In this limit, we obtain an action of the same form as in Ref. \cite{DaPorLie}, as we had anticipated.

Again, we can see  that the coefficients of the contributions that may experience a flip of sign in $v$ vanish if this flip takes place. Therefore, Hilbert subspaces with support on discrete semilattices of step four are left invariant under the action of the Lorentzian Hamiltonian. In view of this, we conclude that the superselection sectors of the Euclidean and the Lorentzian parts of the Hamiltonian constraint coincide. As a consequence, the introduction of the Lorentzian part and its independent regularization and quantization do not alter the superselection sectors of the standard formalism of flat FLRW LQC. Moreover, we have shown that this statement is robust even if we change the operator representation of $1/\bar{\mu}$ and the particular choice of the parameter $\alpha$ for the factor ordering of the powers of the volume operator.

We close the Appendix by remarking that these results can be immediately incorporated in the characterization of the generalized eigenfunctions of the Hamiltonian constraint. In Ref. \cite{DaPorLie}, we concluded that the value of a generalized eigenfunction at a finite point of the semilattice that provides the superselection sector of the volume can be obtained from its values at the two first points. The expression that gives the generalized eigenfunction has two terms, each of which is proportional to the value at either the first or the second point of the semilattice under consideration. We designed a strategy to obtain the respective proportionality factors (which only depend on the superselection sector for the volume $v$ and on the corresponding eigenvalue $\lambda$). This is achieved via the repeated multiplication of a series of functions, that we named $F^{\pm 4}(v)$, $F^{0}_\lambda(v)$, and $F^{-8}(v)$, combined in a way which is determined by the paths that can connect a certain pair of integers through steps of one, two, three, or four units \cite{DaPorLie}. Exactly the same line of reasoning applies to the situation that we are now dealing with. However, the relevant functions, that we now call $\tilde{F}^{\pm 4}_\alpha(v)$, $\tilde{F}^{0}_{\lambda,\alpha}(v)$, and $\tilde{F}^{-8}_\alpha(v)$, are slightly different owing to our modifications in the representation. On the one hand, the functions $f^{(n)}_{\pm}(v)$, used in Ref. \cite{DaPorLie}, must be replaced with the new functions $\tilde{f}^{(n,\alpha)}_{\pm}(v)$. On the other hand, additional terms appear due to the nonvanishing of $\tilde{g}^{(\alpha)}_0(v)$ for arbitrary values of $\alpha$. This affects $\tilde{F}^{\pm 4}_\alpha(v)$ and $\tilde{F}^{0}_{\lambda,\alpha}(v)$, since only the terms proportional to $\ket{v\pm4}$ and $\ket{v}$ receive new contributions in Eq. \eqref{ActionHLtilde}. The final outcome is
\begin{eqnarray}
\tilde{F}^{0}_{\lambda,\alpha}(v) &=&\dfrac{4\gamma^2}{1+\gamma^2}\dfrac{\lambda-\left\{\left[\tilde{f}_{+}^{(2,0)}(v)\right]^2+\left[\tilde{f}_{-}^{(2,0)}(v)\right]^2\right\}}{\tilde{f}_{-}^{(4,\alpha)}(v+8)\tilde{f}_{-}^{(4,\alpha)}(v+4)}+\dfrac{\left[\tilde{f}_{+}^{(4,\alpha)}(v)\right]^2+\left[\tilde{f}_{-}^{(4,\alpha)}(v)\right]^2-\left[\tilde{g}^{(\alpha)}_0(v)\right]^2}{\tilde{f}_{-}^{(4,\alpha)}(v+8)\tilde{f}_{-}^{(4,\alpha)}(v+4)},\\
\tilde{F}^{\pm 4}_\alpha(v)&=& \dfrac{4\gamma^2}{1+\gamma^2}\dfrac{\tilde{f}_{\mp}^{(2,0)}(v\pm 4)\tilde{f}_{\mp}^{(2,0)}(v\pm 2)}{\tilde{f}_{-}^{(4,\alpha)}(v+8)\tilde{f}_{-}^{(4,\alpha)}(v+4)}\pm\dfrac{\tilde{f}_{\mp}^{(4,\alpha)}(v\pm 4)\left[\tilde{g}^{(\alpha)}_0(v)+\tilde{g}^{(\alpha)}_0(v\pm 4)\right]}{\tilde{f}_{-}^{(4,\alpha)}(v+8)\tilde{f}_{-}^{(4,\alpha)}(v+4)},\\
\tilde{F}^{-8}_\alpha(v)&=& -\dfrac{\tilde{f}_{+}^{(4,\alpha)}(v-8)\tilde{f}_{+}^{(4,\alpha)}(v-4)}{\tilde{f}_{-}^{(4,\alpha)}(v+8)\tilde{f}_{-}^{(4,\alpha)}(v+4)}.
\end{eqnarray}
In the limit $\alpha\to 0$, where the usual operator representation is recovered, all new contributions (proportional to $\tilde{g}^{(\alpha)}_0$) vanish. Therefore, in this case the generalized eigenfunctions are only modified  by the replacement $f^{(n)}_{\pm}(v)\rightarrow \tilde{f}^{(n,\alpha)}_{\pm}(v)$. 

\acknowledgments

The authors are thankful to B. Elizaga Navascu\'es for conversations. This work has been supported by Project. No. FIS2017-86497-C2-2-P of MICINN from Spain. The project that gave rise to these results received the support of a fellowship from ``la Caixa'' Foundation (ID 100010434). The fellowship code is LCF/BQ/DR19/11740028.


\begin{thebibliography}{50}


\bibitem{ALQG} A. Ashtekar and J. Lewandowski, Background independent quantum gravity: a status report,  Class. Quantum Grav. \textbf{21}, R53 (2004).
	
\bibitem{Thiem} T. Thiemann, \textit{Modern Canonical Quantum General Relativity} (Cambridge University Press, Cambridge, UK, 2007).


\bibitem{Wald} R.M. Wald, \textit{General Relativity} (Chicago University Press, Chicago, 1984).

\bibitem{Hawking-Ellis} S.W. Hawking and G.F.R. Ellis, \textit{The Large Scale Structure of Space-time} (Cambridge University Press, Cambridge, UK, 1973).

\bibitem{CT} C. Cohen-Tannoudji, B. Diu, and F. Laloe, \textit{Quantum Mechanics} (Wiley, 1977), Vol. 1.
	
\bibitem{Galindo} A. Galindo and P. Pascual, \textit{Quantum Mechanics I} (Springer-Verlag, Berlin, 1990).


\bibitem{Dirac} P.A.M. Dirac, \textit{Lectures on Quantum Mechanics} (Belfer Graduate School Monograph Series, New York, 1964), Vol. 2.


\bibitem{AS} A. Ashtekar and P. Singh, Loop quantum cosmology: a status report, Class. Quantum Grav. \textbf{28}, 213001 (2011).

\bibitem{LQCG} G.A. Mena Marug\'an, A brief introduction to Loop Quantum Cosmology, AIP Conf. Proc. \textbf{1130}, 89  (2009).


\bibitem{APS1} A. Ashtekar, T. Paw\l{}owski, and P. Singh, Quantum nature of the big bang: an analytical and numerical investigation, Phys. Rev. D \textbf{73}, 124038 (2006).
	
\bibitem{APS2} A. Ashtekar, T. Paw{\l}owski, and P. Singh, Quantum nature of the big bang: improved dynamics, Phys. Rev. D \textbf{74}, 084003 (2006).
	
\bibitem{MMO} M. Mart\'{i}n-Benito, G.A. Mena Marug\'an, and J. Olmedo, Further improvements in the understanding of isotropic loop quantum cosmology, Phys. Rev. D \textbf{80}, 104015 (2009).

    
\bibitem{chiou1} D.W. Chiou, Loop quantum cosmology in Bianchi type I models: analytical investigation, Phys. Rev. D \textbf{75}, 024029 (2007).
	
\bibitem{chiou2} D.W. Chiou, Effective dynamics, big bounces, and scaling symmetry in Bianchi type I loop quantum cosmology, Phys. Rev. D \textbf{76}, 124037 (2007).

\bibitem{Bianchii} M. Mart\'in-Benito, G.A. Mena Marug\'an, and T. Paw{\l}owski, Loop quantization of vacuum Bianchi I cosmology, Phys. Rev. D \textbf{78}, 064008 (2008).

\bibitem{AW-E1} A. Ashtekar and E. Wilson-Ewing, Loop quantum cosmology of Bianchi I models, Phys. Rev. D \textbf{79}, 083535 (2009).

\bibitem{Bianchiii} M. Mart\'in-Benito, G.A. Mena Marug\'an, and T. Paw{\l}owski, Physical evolution in loop quantum cosmology: the example of vacuum Bianchi I, Phys. Rev. D \textbf{80}, 084038 (2009).

	
\bibitem{GMM1} M. Mart\'{\i}n-Benito, L.J. Garay, and G.A. Mena Marug\'an, Hybrid quantum Gowdy cosmology: combining loop and Fock quantizations, Phys. Rev. D \textbf{78}, 083516 (2008).
	
\bibitem{GM} G.A. Mena Marug\'an and M. Mart\'in-Benito, Hybrid quantum cosmology: combining loop and Fock quantizations, Int. J. Mod. Phys. A \textbf{24}, 2820 (2009).
    
\bibitem{Gowdy1} L.J. Garay, M. Mart\'in-Benito, and G.A. Mena Marug\'an, Inhomogeneous loop quantum cosmology: hybrid quantization of the Gowdy model, Phys. Rev. D \textbf{82}, 044048 (2010).
    
\bibitem{MMW-E} M. Mart\'in-Benito, G.A. Mena Marug\'an, and E. Wilson-Ewing, Hybrid quantization: from Bianchi I to the Gowdy model, Phys. Rev. D \textbf{82}, 084012 (2010).

    
\bibitem{TT} T. Thiemann, Quantum spin dynamics (QSD), Class. Quantum Grav. \textbf{15}, 839 (1998).
    
\bibitem{ALM} M. Assanioussi, J. Lewandowski, and I. M\"akinen, New scalar constraint operator for loop quantum gravity, Phys. Rev. D \textbf{92}, 044042 (2015).
    
\bibitem{AAL} E. Alesci, M. Assanioussi, and J. Lewandowski, Curvature operator for LQG, Phys. Rev. D \textbf{89}, 124017 (2014).
	
    
\bibitem{Engle} J. Engle, Relating loop quantum cosmology to loop quantum gravity: symmetric sectors and embeddings, Class. Quantum Grav. \textbf{24}, 5777 (2007).
    
\bibitem{BK} J. Brunnemann and T.A. Koslowski, Symmetry reduction of loop quantum gravity, Class. Quantum Grav. \textbf{28}, 245014 (2011).

\bibitem{Engle2} J. Engle, Embedding loop quantum cosmology without piecewise linearity, Class. Quantum Grav. \textbf{30}, 085001 (2013).

\bibitem{Paw} T. Paw{\l}owski, Observations on interfacing loop quantum gravity with cosmology, Phys. Rev. D \textbf{92}, 124020 (2015).

\bibitem{BEHM} C. Beetle, J. Engle, M.E. Hogan, and P. Mendo\c{c}a, Diffeomorphism invariant cosmological sector in loop quantum gravity, Class. Quantum Grav. \textbf{34}, 225009 (2017). 
    
    
\bibitem{DL1} A. Dapor and K. Liegener, Cosmological effective Hamiltonian from full loop quantum gravity dynamics, Phys. Lett. B \textbf{785}, 506 (2018).

\bibitem{DL2} M. Assanioussi, A. Dapor, K. Liegener, and T. Paw{\l}owski, Emergent de Sitter epoch of the quantum cosmos from loop quantum cosmology, Phys. Rev. Lett. \textbf{121}, 081303 (2018).

\bibitem{DL3} M. Assanioussi, A. Dapor, K. Liegener, and T. Paw{\l}owski, Emergent de Sitter epoch of the loop quantum cosmos: a detailed analysis, Phys. Rev. D \textbf{100}, 084003 (2019).
    
\bibitem{YDM} J. Yang, Y. Ding, and Y. Ma, Alternative quantization of the Hamiltonian in loop quantum cosmology, Phys. Lett. B \textbf{682}, 1 (2009).    
    
\bibitem{Paramc1} B.-F. Li, P. Singh, and A. Wang, Towards cosmological dynamics from loop quantum gravity, Phys. Rev. D \textbf{97},  084029 (2018).
    
\bibitem{Paramc2} B.-F. Li, P. Singh, and A. Wang, Qualitative dynamics and inflationary attractors in loop cosmology, Phys. Rev. D \textbf{98}, 066016 (2018).
    
\bibitem{genericness} B.-F. Li, P. Singh, and A. Wang, Genericness of pre-inflationary dynamics and probability of slow-roll inflation in modified loop quantum cosmologies, Phys. Rev. D \textbf{100}, 063513 (2019).
    
\bibitem{Agullo} I. Agullo, Primordial power spectrum from the Dapor-Liegener model of loop quantum cosmology, Gen. Relativ. Gravit. \textbf{50}, 91 (2018).

\bibitem{Haro} J. de Haro, The Dapor-Liegener model of loop quantum cosmology: a dynamical analysis, Eur. Phys. J. C \textbf{78}, 926 (2018).


\bibitem{DaPorLie} A. Garc\'ia-Quismondo and G.A. Mena Marug\'an, Mart\'in-Benito---Mena Marug\'an---Olmedo prescription for the Dapor-Liegener model of loop quantum cosmology, Phys. Rev. D \textbf{99}, 083505 (2019).


\bibitem{gaugeinvariant} K. Liegener and P. Singh, Gauge invariant bounce from quantum geometry, arXiv:1906.02759.
    
\bibitem{CSaction} J. Yang, C. Zhang, and Y. Ma, Loop quantum cosmology from an alternative Hamiltonian, Phys. Rev. D \textbf{100}, 064026 (2019).

\bibitem{kramer} D. Kramer, H. Stephani, M. MacCallum, and E. Herlt, \textit{Exact Solutions of Einstein Field Equations} (Cambridge University Press, Cambridge, UK, 1980).


\end{thebibliography}
\end{document}